\documentclass[aip,apl,reprint,showpacs,superscriptaddress,groupedaddress]{revtex4-1}  % for review and submission
\usepackage{graphicx}  % needed for figures
\usepackage{dcolumn}   % needed for some tables
\usepackage{bm}        % for math
\usepackage{amsmath,amssymb}   % for math
\usepackage{color}
\usepackage{pdfpages}
\usepackage{pgffor}

\makeatletter
\AtBeginDocument{\let\LS@rot\@undefined}
\makeatother

\begin{document}

\title{Graphene Nanoribbons on Hexagonal Boron Nitride: Deposition and Transport Characterization}
\author{Tobias Preis} \affiliation{Institute of Experimental and Applied Physics, University of Regensburg, D-93040 Regensburg, Germany}
\author{Christian Kick} \affiliation{Institute of Experimental and Applied Physics, University of Regensburg, D-93040 Regensburg, Germany}
\author{Andreas Lex} \affiliation{Institute of Experimental and Applied Physics, University of Regensburg, D-93040 Regensburg, Germany}
\author{Akimitsu Narita} \affiliation{Max Planck Institute for Polymer Research, Ackermannweg 10, D-55128 Mainz, Germany}
%\author{Yunbin Hu} \affiliation{Max Planck Institute for Polymer Research, Ackermannweg 10, D-55128 Mainz, Germany}
\author{Yunbin Hu} \altaffiliation[Current address:]{Department of Organic and Polymer Chemistry, College of Chemistry and Chemical Engineering, Central South University, Changsha, Hunan 410083, China}
\affiliation{Max Planck Institute for Polymer Research, Ackermannweg 10, D-55128 Mainz, Germany}
\author{Kenji Watanabe} \affiliation{National Institute for Materials Science, 1-1 Namiki, Tsukuba 305-0044, Japan}
\author{Takashi Taniguchi} \affiliation{National Institute for Materials Science, 1-1 Namiki, Tsukuba 305-0044, Japan}
\author{Klaus M\"ullen} \affiliation{Max Planck Institute for Polymer Research, Ackermannweg 10, D-55128 Mainz, Germany}
\author{Dieter Weiss} \affiliation{Institute of Experimental and Applied Physics, University of Regensburg, D-93040 Regensburg, Germany}
\author{Jonathan Eroms} \email{jonathan.eroms@ur.de}\affiliation{Institute of Experimental and Applied Physics, University of Regensburg, D-93040 Regensburg, Germany}

\date{\today}

\begin{abstract}
Chemically synthesized ``cove''-type graphene nanoribbons (cGNRs) of different widths were brought into dispersion and drop-cast onto exfoliated hexagonal boron nitride (hBN) on a Si/SiO$_2$ chip. With AFM we observed that the cGNRs form ordered domains aligned along the crystallographic axes of the hBN. Using electron beam lithography and metallization, we contacted the cGNRs with NiCr/Au, or Pd contacts and measured their $I$-$V$-characteristics. The transport through the ribbons was dominated by the Schottky behavior of the contacts between the metal and the ribbon.

\end{abstract}

\maketitle

%\section{\label{sec:level1}First-level heading}
% sections are not used for PRL papers
Confining graphene in one dimension yields graphene nanoribbons (GNRs), which have great potential for application in semiconductor technology.
Depending on their width and edge configuration, GNRs can have a bandgap that, e.g., allows turning on and off the current flow in the GNR which is needed to design transistors.\cite{Schwierz2010, Nakada1996, Han2007}
Initially, GNRs were obtained by top-down methods like etching structured graphene sheets\cite{Han2007} or unzipping carbon nanotubes\cite{Kosynkin2009, Jiao2009}. However, these ribbons had rough edges which limit carrier transport and thus the usability of the GNRs in devices.\cite{Han2010gnrs, Evaldsson2008} {Epitaxially grown GNRs on SiC show ballistic transport\cite{Baringhaus2014}, and GNRs were prepared inside etched trenches in hexagonal boron nitride (hBN)\cite{Chen2017}, but both methods again lack atomic precision.}
Advances in solution chemistry opened up new routes to obtain GNRs by atomically precise bottom-up synthesis. There, GNRs are obtained by reactions of precursor molecules on catalytic metal surfaces.\cite{Cai2010} However, it is not possible to measure the GNRs' transport properties on a metal surface.
%Therefore the GNRs have to be transferred onto insulating substrates (typically Si/SiO$_2$) which usually is connected with etchants that contaminate the GNRs.\cite{Bennett2013} Furthermore, SiO$_2$ has a rough surface which negatively influences the carrier transport. The mobility of graphene was shown to increase considerably when the graphene flakes were put on hexagonal boron nitride (hBN) instead of Si/SiO$_2$.\cite{Dean2010}
One way to overcome this issue is to transfer the GNRs to insulating substrates (typically Si/SiO$_2$).\cite{Bennett2013, Chen2016, Llinas2017} The disadvantage here is that the transfer process usually involves etchants that contaminate the GNRs. Instead, solution-processable GNRs can be employed.\cite{Narita2014natchem, Abbas2014, Zschieschang2015, Fantuzzi2016} Here, GNR powder is dispersed in a solvent and then drop-cast onto an arbitrary surface. In previous experiments, SiO$_2$ substrates were used, whose rough surface and charged impurities negatively influence carrier transport. The mobility of extended graphene was shown to increase considerably when graphene  was placed onto hexagonal boron nitride (hBN) instead of Si/SiO$_2$.\cite{Dean2010} {For etched GNRs on hBN, on the other hand, sample properties did not improve due to disorder introduced by plasma etching\cite{Bischoff2012}}. However, deposition and device fabrication of {bottom-up-synthesized} GNRs on hBN has not been reported. In this work we demonstrate the deposition of GNRs onto the atomically flat surface of exfoliated hBN, showing a unique self-assembly behavior with domains of parallely aligned GNRs over tens to hundreds of nm.  We further discuss the fabrication and characteristics of GNR-based FET devices on hBN. 

\begin{figure}
\includegraphics[width=0.5\textwidth]{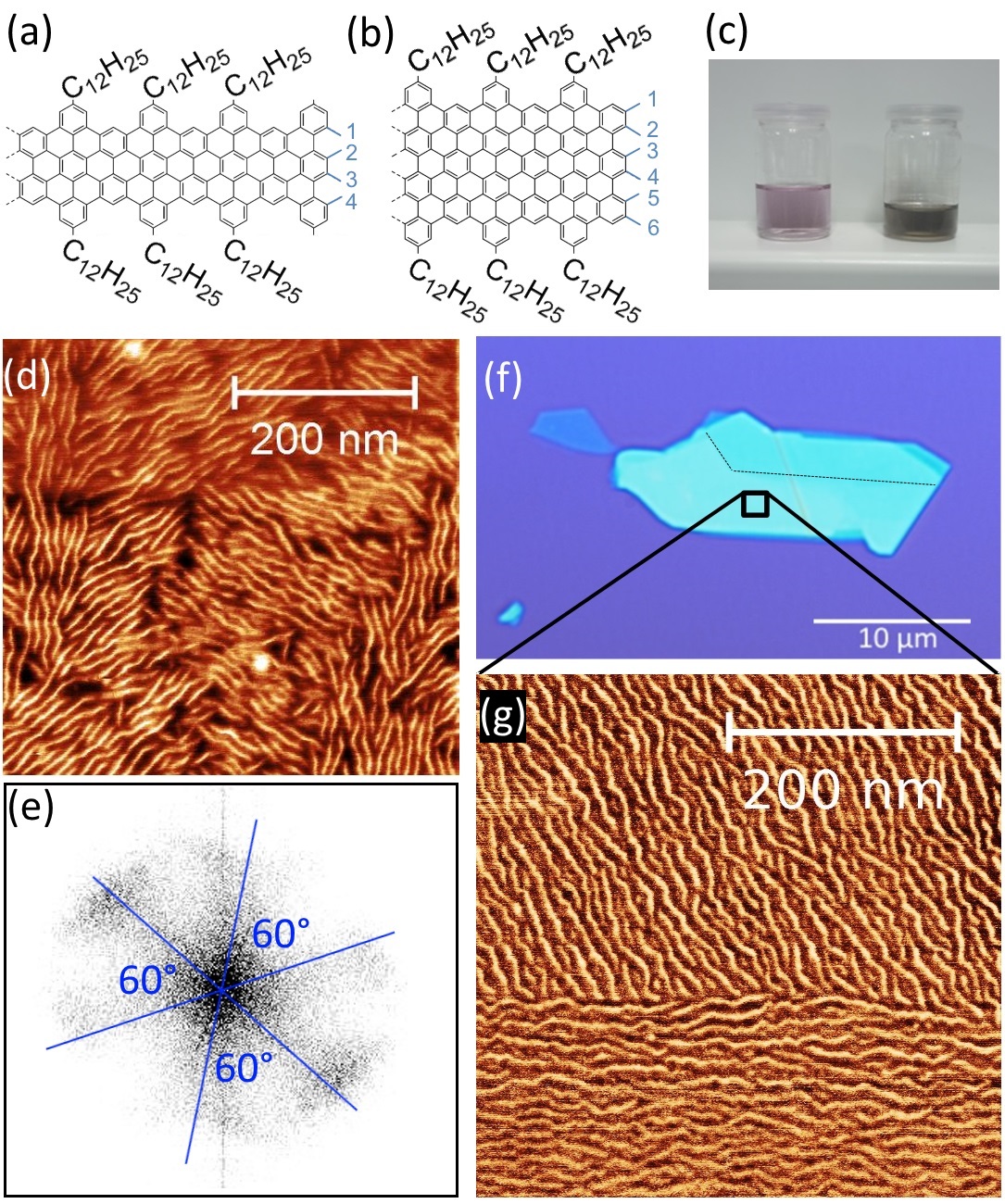}
\caption{\label{fig:img1}cGNRs on hBN. {(a) -- (b) Chemical structure of 4- and 6-cGNRs. (c) Dispersion obtained after sonicating cGNR powder in THF (4-cGNR, left) and chlorobenzene (6-cGNR, right). (d) AFM phase image of 4-cGNRs on hBN. (e) 2D-FFT of (d) showing the preferential directions of the cGNRs. (f) Optical microscope image of an exfoliated hBN flake on Si/SiO$_2$. (g) AFM phase image of 6-cGNRs (indicated area in (f)).}}
\end{figure}

We investigated solution-processable ``cove''-type GNRs (cGNRs) of different widths\cite{Narita2014natchem,Hu2018} (4 and 6 carbon dimers, see Fig. \ref{fig:img1} (a), (b)). These cGNRs were predicted to have a band gap between 1.5\,eV and 2.0\,eV.\cite{Osella2012,Ivanov2017} The alkyl-side chains, which are attached for better solubility, were shown to have no substantial effect on the electronic structure.\cite{Villegas2014} To fabricate devices, cGNR powder (see our previous reports for the syntheses of the 4-cGNRs\cite{Narita2014natchem} and 6-cGNRs\cite{Hu2018}) was put in tetrahydrofuran (THF) for the 4-cGNRs, or in chlorobenzene for the 6-cGNRs, respectively. Hereafter, the mixture was sonicated for at least 1\,hour, until the powder was mostly dispersed and the dispersion turned violet (gray) for the 4-cGNRs (6-cGNRs), as can be seen in Fig. \ref{fig:img1} (c).

Subsequently, we exfoliated hBN on a Si/SiO$_2$ chip,  drop-cast the cGNR dispersion onto the chip, and evaporated the solvent by placing the chip  onto a hot plate.
% hBN as a substrate was shown to substantially improve the carrier mobility of graphene compared to graphene on SiO$_2$, so the question arises whether it has a similar effect on cGNRs.
Afterwards, we investigated the flakes with an atomic force microscope (AFM). Fig. \ref{fig:img1} (d) and (h) show AFM phase images of 4- and 6-cGNRs, respectively, on hBN flakes. {In previous studies on SiO$_2$, it was found that cGNRs only absorb on carefully functionalized substrates\cite{Abbas2014,Zschieschang2015}. Similarly, we did not observe cGNR adsorption on the untreated SiO$_2$ surface on our samples. In contrast,} we found that cGNRs adsorb readily to the atomically flat hBN surfaces and form well-ordered domains with domain sizes ranging from 60\,nm to over 1\,$\mu$m and ribbon lengths of up to 350\,nm. {Length distributions for both cGNR types can be found in the Supplemental Material.} The situation is similar to adsorption on highly-oriented pyrolitic graphite (HOPG) surfaces \cite{Narita2014natchem}, but with an important difference. While cGNRs on HOPG form arrays of straight ribbons, here we find that individual GNRs have a wiggled structure. Although we cannot offer a clear explanation for this observation, a possible reason could be the slight lattice mismatch between hBN and the graphene backbone of the cGNRs. This was shown in molecular dynamics simulations to lead to lateral buckling and snake-like motion of GNRs \cite{Ouyang2018}. Alternatively, non-planar adsorption of the alkyl-side chains on the hBN could play a role. 
{It should be further mentioned that the cGNRs seem to be very mobile on the hBN surface. cGNR covered hBN flakes were annealed to 450 $^\circ$C and then re-investigated with AFM. Before, the hBN flake was homogeneously covered with cGNRs, after the annealing the cGNRs seem to have formed agglomerates (see Supplementary Material) and the bare hBN flake is visible again. The superlubricity of armchair GNRs on a gold substrate was  already reported in UHV experiments.\cite{Kawai2016}}
{Finally, }we note that, although we assume that the cGNRs form monolayers on the hBN flakes, given the $z$-resolution of our AFM we cannot rule out that more than one layer of cGNRs is adsorbed on the hBN. The ordered domains of cGNRs on hBN are found to be rotated by 60$^\circ$ with respect to each other. The angles between the domains become especially clear when we plot a two dimensional Fast Fourier Transformation (2D-FFT) of the AFM phase image (Fig. \ref{fig:img1} (e)). The broadening of the 2D-FFT is mainly due to the wiggled structure of the cGNRs. 

Fig. \ref{fig:img1} (f) shows an optical microscope image of an exfoliated hBN flake on a Si/SiO$_2$ chip before deposition of 6-cGNRs.
When exfoliating hBN onto SiO$_2$, it often cleaves along its crystallographic axes.
Considering the hexagonal lattice structure of hBN consisting of alternating B and N atoms\cite{Taniguchi2007}, this yields cleaving angles in multiples of 30$^\circ$.
Two of those axes are indicated with black dashed lines in Fig. \ref{fig:img1} (f).
Fig. \ref{fig:img1} (g) is an AFM phase image of the area enclosed by the black square in Fig. \ref{fig:img1} (f) after drop-casting the 6-cGNRs onto the chip.
When comparing the orientation of the cGNRs in Fig. \ref{fig:img1} (g) and the edges of the hBN flake in Fig. \ref{fig:img1} (f), it becomes apparent that the cGNR domains are aligned along the crystallographic axes of the hBN.

\begin{figure}
\includegraphics[width=0.5\textwidth]{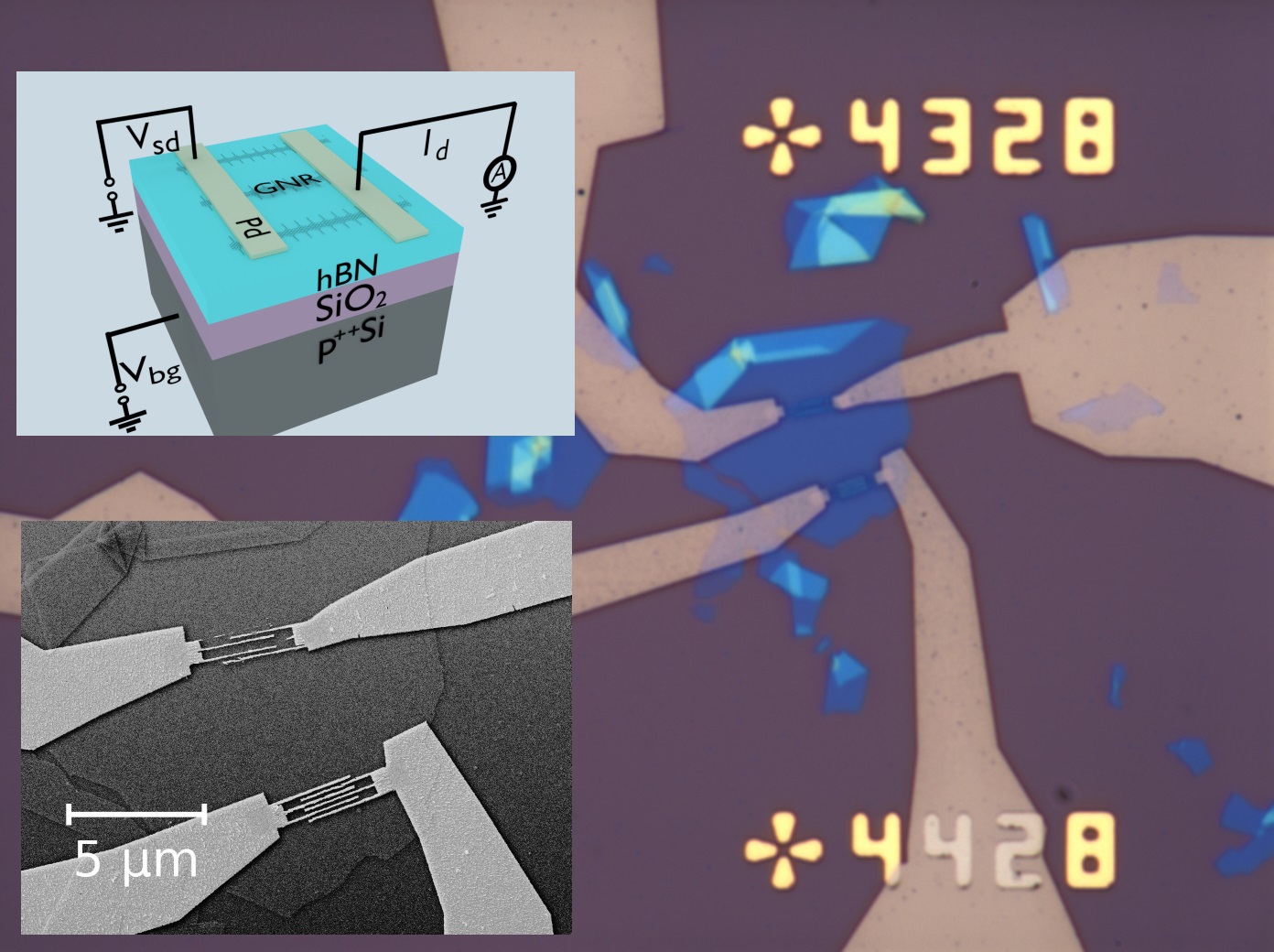}
\caption{\label{fig:img2}Sample design. {Optical microscope image of an hBN flake on Si/SiO$_2$ with contacts reaching the cGNRs on the flake. (Distance between markers: 50\,$\mu$m.) Lower inset: Magnified SEM image of the contact combs on the flake. Upper inset: Schematics of the measurement setup.}}
\end{figure}

Next, we contacted the cGNRs by performing electron beam lithography and evaporating metals (thermally and e-beam). NiCr/Au or Pd\cite{Xia2011} served as contact materials. {Pd was deposited without any adhesion layer, therefore care had to be taken during lift-off not to damage the fine metal structures.}
We contacted multiple cGNRs at once, using interdigitated comb-like structures (see lower inset of Fig. \ref{fig:img2}). The orientation of the contact combs was chosen in such a way that the contacts were perpendicular to some of the cGNR domains.
Fig. \ref{fig:img2} shows an optical microscope image of one device with 15\,nm thick Pd contacts.

The schematics of the measurement setup is sketched in the upper inset of Fig. \ref{fig:img2}. With a probe station at ambient conditions, a source drain voltage $V_{sd}$ was applied across the ribbons and the drain current $I_d$ was measured. The heavily p-doped Si substrate could be used as a back gate electrode by applying a back gate voltage $V_{bg}$. The SiO$_2$ layer was 285\,nm thick, the thickness of the used hBN flakes varied between 10-40\,nm and the spacing of the contacts between 70-120\,nm.

\begin{figure}
\includegraphics[width=0.5\textwidth]{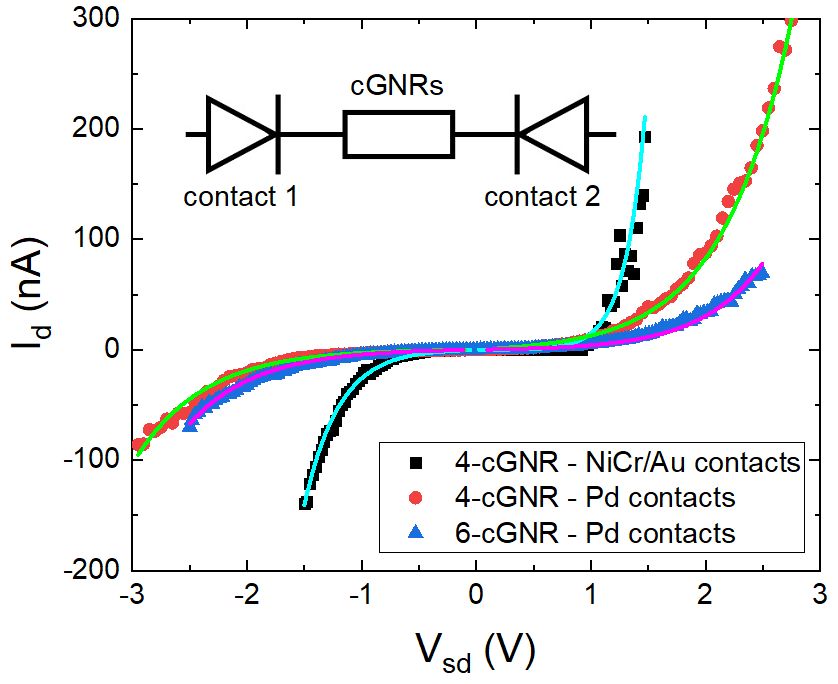}
\caption{\label{fig:img3}$I$-$V$-curves. {$I$-$V$-measurements of 4- and 6-cGNRs contacted with NiCr/Au and Pd (dots) and fits for the negative and positive voltage regions (lines). Inset: Schematics for two Schottky contacts connected back-to-back over a cGNR.}}
\end{figure}

Fig. \ref{fig:img3} shows $I$-$V$-measurements of 4- and 6-cGNRs contacted with NiCr/Au and Pd. The NiCr/Au contacted 4-cGNRs (black squares) show a current onset for the lowest source drain voltages. We note that all curves are asymmetric with respect to $V_{sd}=0$. {Possible sources for this asymmetry could be slightly different work functions of the electrodes due to contamination\cite{Perello2011}, different contact areas\cite{Zhang2007}, or the fact that the bias voltage is not applied symmetrically, but with one terminal grounded \cite{Appenzeller2002}.}

\begin{figure}
	\includegraphics[width=0.5\textwidth]{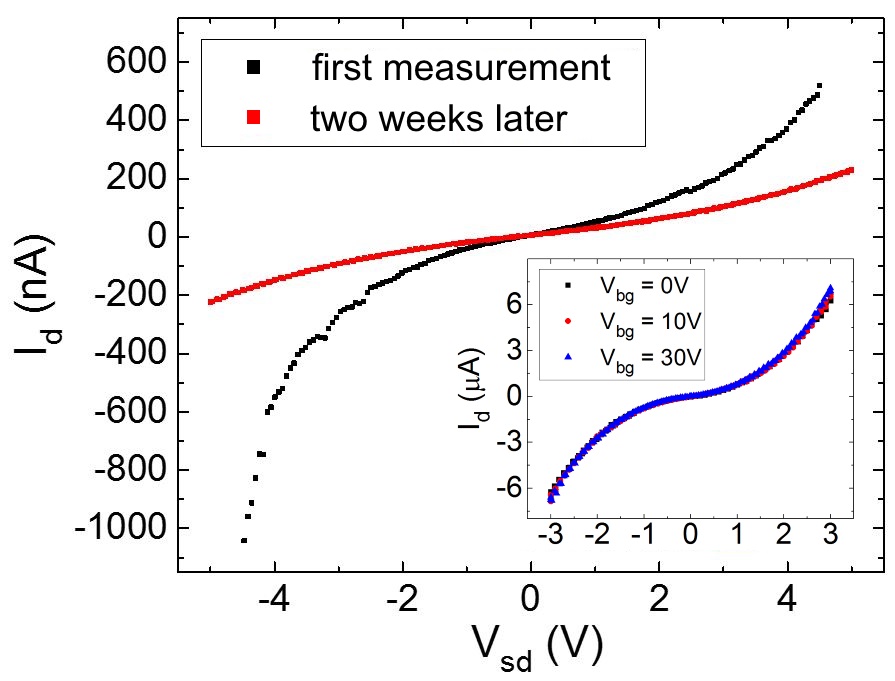}
	\caption{\label{fig:img4}Further measurements of 6-cGNRs. {$I$-$V$-curves have been taken directly after device fabrication and two weeks later. The inset shows $I$-$V$-curves of a different sample at various back gate voltages.}}
\end{figure}

The Pd contacted 4- and 6-cGNRs look very similar on the negative voltage side and differ only slightly on the positive side. This is quite surprising since their different band gap should be reflected in the $I$-$V$-curves. Here we assume that the Fermi level is always situated in the gap, as the curves can be described by a Schottky characteristic (see below).
Further, when measuring contact pairs of further devices (of the same kind of cGNR and the same contact metal) the shape of the $I$-$V$-curves deviated from the curves shown in Fig. \ref{fig:img3} and also the drain currents varied even by orders of magnitudes (see, e.g., inset of Fig. \ref{fig:img4}).
Taking all these effects into account, it seems likely that the measurements are dominated by the Schottky behavior of the contacts between the semiconducting cGNRs and the metallic electrodes.
The transition from the metal to the cGNR and back to the metal can be seen as two back-to-back connected Schottky diodes\cite{Chiquito2012, Sze2007} with a resistor (one or many cGNRs) in between (see inset of Fig. \ref{fig:img3}).
Since the detected current is limited by the current leaking through the Schottky diode in reverse direction, we fitted the Schottky barriers separately for negative and positive voltage regions (for the reverse direction of the Schottky diode). {The slight difference in barrier height for positive and negative bias was included in the error margin.}
{We found that the current density $J$ of our data is best described by the thermionic field emission model:\cite{Padovani1966,Zhang2007,Liu2008,Perello2011}
\begin{eqnarray}
J=&&\frac{A T\sqrt{\pi q E_{00}}}{k}\exp\left(-\frac{\Phi}{qE_0}\right)\exp\left[ V\left( \frac{q}{kT}-\frac{1}{E_0} \right) \right]\nonumber\\
&&\times\sqrt{q(V-\zeta) + \frac{\Phi}{\cosh^2(qE_{00}/kT)}}
\end{eqnarray}
Here, $A$ is the Richardson constant, $T$ the temperature, $\Phi$ the height of the Schottky barrier, $k$ the Boltzmann constant, $q$ the elementary charge, $V>0$ the voltage applied across the barrier, and $\zeta$ the distance between the band edge and the Fermi level. $E_{00}$ describes the shape of the barrier\cite{Padovani1966} and $E_0=E_{00}\coth(qE_{00}/kT)$. For the contact area, we take the cross-sectional area of one nanoribbon, but include the unknown number of parallel ribbons, and further sources of uncertainty, in the error margin of $\Phi$ (for more details, see the Supplemental Material). }
\begin{table}
	\caption{\label{tab:fit}{Fitting results for the graphs in Fig. \ref{fig:img3}.}}
	\begin{ruledtabular}
		\begin{tabular}{rccc}
	Sample&$\Phi$&$E_{00}$&$\zeta$\\
			\hline
			NiCr/Au 4-cGNR & $(190\pm180)$ meV  & 16 mV&0.15 V\\ 			
			Pd 4-cGNR & $(130\pm180)$ meV &  9 mV&0.2 V\\
			Pd 6-cGNR& $(150\pm150)$ meV  &  9 mV&0.2 V\\
		\end{tabular}
	\end{ruledtabular}
\end{table}
{The fitting results are summarized in Table \ref{tab:fit}. NiCr/Au yields a higher barrier than the Pd contacts.} 
Furthermore, Pd contacts worked more reliably than NiCr/Au contacts which is why no data are shown for NiCr/Au contacted 6-cGNRs.
Taking these facts into account, Pd seems to be the better contact material.
The above numbers, however, have to be considered with care. Strictly speaking, the Richardson constant $A$ is only valid for the free electron mass, but has to be modified using the electron effective mass \cite{Crowell1965,Sze2007}, which is unknown for our contact configuration. {We therefore used the free electron mass.}
 Also, the alkyl side chains could partially overlap with the cGNRs,\cite{Narita2014natchem} leading to a further increase of the contact resistance.
{Both uncertainties are absorbed in the error of the barrier height $\Phi$. The obtained values for $\Phi$ are in line with earlier results on carbon nanotubes.\cite{Svensson2011}
}

Additionally we studied the stability of our fabricated devices in air. Fig. \ref{fig:img4} shows Pd contacted 6-cGNRs measured directly after fabrication (black squares) and two weeks later (red squares). The decreased drain current could be due to degradation of the cGNRs, {degradation of the contacts,} or contamination of the cGNRs by particles in the air.

Finally, we studied the gate response of our devices. As can be seen in the inset of Fig. \ref{fig:img4}, the devices showed (almost) no back gate dependence which could be due to Fermi level pinning at the Schottky contacts. Also, we note that the distance to the Si back gate is much larger than the separation of the metal electrodes. Therefore, we expect the gate coupling to be greatly reduced due to screening. 

{In previous studies, GNRs prepared from molecular precursors by either surface synthesis\cite{Llinas2017} or in solution\cite{Abbas2014,Zschieschang2015}, were deposited on oxidized silicon. When depositing 4-cGNRs dispersed in organic solvents or water onto SiO$_2$, careful surface functionalization was necessary\cite{Abbas2014,Konnerth2015,Zschieschang2015}. In this work, on the other hand, cGNRs adsorb readily to the hBN surface. We found dense arrays of cGNRs on almost every hBN flake studied. This presents a clear advantage over the previous preparation method. In addition, hBN was shown to be a more suitable substrate for high-quality graphene devices. With respect to the contact transparency, we note that previous experiments on 4-cGNRs prepared from organic solvents and surface synthesized armchair GNRs found a device current in the 1 to 1000 nA range for a source-drain bias of 1 V, in line with our observations.\cite{Abbas2014,Fantuzzi2016,Llinas2017} This also holds for 6-cGNRs which were not previously used in transport devices. In the case of cGNRs prepared from aqueous solution, Zschieschang {\em et al.} obtained a drain current of tens of $\mu$A. However, their devices showed signs of agglomeration, leading to a greatly reduced band gap, which could explain the lower Schottky barrier.\cite{Zschieschang2015}}

In summary, we dispersed chemically synthesized 4- and 6-cGNRs in THF or cholorobenzene. The dispersion was drop-cast onto exfoliated hBN. cGNRs adsorb readily to the flat hBN surfaces and form ordered domains aligned along the crystallographic axes of the hBN, showing the potential of hBN as a substrate for GNR-based devices. We contacted the cGNRs with NiCr/Au, or Pd contacts. The $I$-$V$-characteristics of the devices are dominated by the Schottky behavior of the contacts between metal and ribbon. {Therefore, better contacts, such as edge-type contacts \cite{Matsuda2010,Wang1Dcontacts,Huang20150D} and local gates are called for, which are technologically more demanding.}

\section*{Supplementary Material}

{See Supplementary Material for an AFM image comparing the cGNR coverage on SiO$_2$ and hBN, for the effect of annealing, {for the length distriution, for details on fitting the $I$-$V$-curves} and for an exemplary $R_{sd}$ vs $V_{bg}$ curve, showing no clear gate dependence.}

\begin{acknowledgments}
Financial support by the Deutsche Forschungsgemeinschaft (DFG) within the programs SFB 689, GRK 1570 and SPP1459 and by the Max Planck Society for the synthesis of cGNRs is gratefully acknowledged.  Growth of hexagonal boron nitride crystals was supported by the Elemental Strategy Initiative conducted by the MEXT, Japan and JSPS KAKENHI Grant Numbers JP15K21722. 
\end{acknowledgments}

%\bibliography{Literatur}

\begin{thebibliography}{40}%
	\makeatletter
	\providecommand \@ifxundefined [1]{%
		\@ifx{#1\undefined}
	}%
	\providecommand \@ifnum [1]{%
		\ifnum #1\expandafter \@firstoftwo
		\else \expandafter \@secondoftwo
		\fi
	}%
	\providecommand \@ifx [1]{%
		\ifx #1\expandafter \@firstoftwo
		\else \expandafter \@secondoftwo
		\fi
	}%
	\providecommand \natexlab [1]{#1}%
	\providecommand \enquote  [1]{``#1''}%
	\providecommand \bibnamefont  [1]{#1}%
	\providecommand \bibfnamefont [1]{#1}%
	\providecommand \citenamefont [1]{#1}%
	\providecommand \href@noop [0]{\@secondoftwo}%
	\providecommand \href [0]{\begingroup \@sanitize@url \@href}%
	\providecommand \@href[1]{\@@startlink{#1}\@@href}%
	\providecommand \@@href[1]{\endgroup#1\@@endlink}%
	\providecommand \@sanitize@url [0]{\catcode `\\12\catcode `\$12\catcode
		`\&12\catcode `\#12\catcode `\^12\catcode `\_12\catcode `\%12\relax}%
	\providecommand \@@startlink[1]{}%
	\providecommand \@@endlink[0]{}%
	\providecommand \url  [0]{\begingroup\@sanitize@url \@url }%
	\providecommand \@url [1]{\endgroup\@href {#1}{\urlprefix }}%
	\providecommand \urlprefix  [0]{URL }%
	\providecommand \Eprint [0]{\href }%
	\providecommand \doibase [0]{http://dx.doi.org/}%
	\providecommand \selectlanguage [0]{\@gobble}%
	\providecommand \bibinfo  [0]{\@secondoftwo}%
	\providecommand \bibfield  [0]{\@secondoftwo}%
	\providecommand \translation [1]{[#1]}%
	\providecommand \BibitemOpen [0]{}%
	\providecommand \bibitemStop [0]{}%
	\providecommand \bibitemNoStop [0]{.\EOS\space}%
	\providecommand \EOS [0]{\spacefactor3000\relax}%
	\providecommand \BibitemShut  [1]{\csname bibitem#1\endcsname}%
	\let\auto@bib@innerbib\@empty
	%</preamble>
	\bibitem [{\citenamefont {Schwierz}(2010)}]{Schwierz2010}%
	\BibitemOpen
	\bibfield  {author} {\bibinfo {author} {\bibfnamefont {F.}~\bibnamefont
			{Schwierz}},\ }\href {\doibase doi:10.1038/nnano.2010.89} {\bibfield
		{journal} {\bibinfo  {journal} {Nature Nanotechnology}\ }\textbf {\bibinfo
			{volume} {5}},\ \bibinfo {pages} {487} (\bibinfo {year} {2010})}\BibitemShut
	{NoStop}%
	\bibitem [{\citenamefont {Nakada}\ \emph {et~al.}(1996)\citenamefont {Nakada},
		\citenamefont {Fujita}, \citenamefont {Dresselhaus},\ and\ \citenamefont
		{Dresselhaus}}]{Nakada1996}%
	\BibitemOpen
	\bibfield  {author} {\bibinfo {author} {\bibfnamefont {K.}~\bibnamefont
			{Nakada}}, \bibinfo {author} {\bibfnamefont {M.}~\bibnamefont {Fujita}},
		\bibinfo {author} {\bibfnamefont {G.}~\bibnamefont {Dresselhaus}}, \ and\
		\bibinfo {author} {\bibfnamefont {M.~S.}\ \bibnamefont {Dresselhaus}},\
	}\href {\doibase https://doi.org/10.1103/PhysRevB.54.17954} {\bibfield
		{journal} {\bibinfo  {journal} {Physical Review B}\ }\textbf {\bibinfo
			{volume} {54}},\ \bibinfo {pages} {17954} (\bibinfo {year}
		{1996})}\BibitemShut {NoStop}%
	\bibitem [{\citenamefont {Han}\ \emph {et~al.}(2007)\citenamefont {Han},
		\citenamefont {\"Ozyilmaz}, \citenamefont {Zhang},\ and\ \citenamefont
		{Kim}}]{Han2007}%
	\BibitemOpen
	\bibfield  {author} {\bibinfo {author} {\bibfnamefont {M.~Y.}\ \bibnamefont
			{Han}}, \bibinfo {author} {\bibfnamefont {B.}~\bibnamefont {\"Ozyilmaz}},
		\bibinfo {author} {\bibfnamefont {Y.}~\bibnamefont {Zhang}}, \ and\ \bibinfo
		{author} {\bibfnamefont {P.}~\bibnamefont {Kim}},\ }\href {\doibase
		10.1103/PhysRevLett.98.206805} {\bibfield  {journal} {\bibinfo  {journal}
			{Physical Review Letters}\ }\textbf {\bibinfo {volume} {98}},\ \bibinfo
		{pages} {206805} (\bibinfo {year} {2007})}\BibitemShut {NoStop}%
	\bibitem [{\citenamefont {Kosynkin}\ \emph {et~al.}(2009)\citenamefont
		{Kosynkin}, \citenamefont {Higginbotham}, \citenamefont {Sinitskii},
		\citenamefont {Lomeda}, \citenamefont {Dimiev}, \citenamefont {Price},\ and\
		\citenamefont {Tour}}]{Kosynkin2009}%
	\BibitemOpen
	\bibfield  {author} {\bibinfo {author} {\bibfnamefont {D.~V.}\ \bibnamefont
			{Kosynkin}}, \bibinfo {author} {\bibfnamefont {A.~L.}\ \bibnamefont
			{Higginbotham}}, \bibinfo {author} {\bibfnamefont {A.}~\bibnamefont
			{Sinitskii}}, \bibinfo {author} {\bibfnamefont {J.~R.}\ \bibnamefont
			{Lomeda}}, \bibinfo {author} {\bibfnamefont {A.}~\bibnamefont {Dimiev}},
		\bibinfo {author} {\bibfnamefont {B.~K.}\ \bibnamefont {Price}}, \ and\
		\bibinfo {author} {\bibfnamefont {J.~M.}\ \bibnamefont {Tour}},\ }\href
	{http://dx.doi.org/10.1038/nature07872} {\bibfield  {journal} {\bibinfo
			{journal} {Nature}\ }\textbf {\bibinfo {volume} {458}},\ \bibinfo {pages}
		{872} (\bibinfo {year} {2009})}\BibitemShut {NoStop}%
	\bibitem [{\citenamefont {Jiao}\ \emph {et~al.}(2009)\citenamefont {Jiao},
		\citenamefont {Zhang}, \citenamefont {Wang}, \citenamefont {Diankov},\ and\
		\citenamefont {Dai}}]{Jiao2009}%
	\BibitemOpen
	\bibfield  {author} {\bibinfo {author} {\bibfnamefont {L.}~\bibnamefont
			{Jiao}}, \bibinfo {author} {\bibfnamefont {L.}~\bibnamefont {Zhang}},
		\bibinfo {author} {\bibfnamefont {X.}~\bibnamefont {Wang}}, \bibinfo {author}
		{\bibfnamefont {G.}~\bibnamefont {Diankov}}, \ and\ \bibinfo {author}
		{\bibfnamefont {H.}~\bibnamefont {Dai}},\ }\href
	{http://dx.doi.org/10.1038/nature07919} {\bibfield  {journal} {\bibinfo
			{journal} {Nature}\ }\textbf {\bibinfo {volume} {458}},\ \bibinfo {pages}
		{877} (\bibinfo {year} {2009})}\BibitemShut {NoStop}%
	\bibitem [{\citenamefont {Han}\ \emph {et~al.}(2010)\citenamefont {Han},
		\citenamefont {Brant},\ and\ \citenamefont {Kim}}]{Han2010gnrs}%
	\BibitemOpen
	\bibfield  {author} {\bibinfo {author} {\bibfnamefont {M.~Y.}\ \bibnamefont
			{Han}}, \bibinfo {author} {\bibfnamefont {J.~C.}\ \bibnamefont {Brant}}, \
		and\ \bibinfo {author} {\bibfnamefont {P.}~\bibnamefont {Kim}},\ }\href
	{https://link.aps.org/doi/10.1103/PhysRevLett.104.056801} {\bibfield
		{journal} {\bibinfo  {journal} {Physical Review Letters}\ }\textbf {\bibinfo
			{volume} {104}},\ \bibinfo {pages} {056801} (\bibinfo {year}
		{2010})}\BibitemShut {NoStop}%
	\bibitem [{\citenamefont {Evaldsson}\ \emph {et~al.}(2008)\citenamefont
		{Evaldsson}, \citenamefont {Zozoulenko}, \citenamefont {Xu},\ and\
		\citenamefont {Heinzel}}]{Evaldsson2008}%
	\BibitemOpen
	\bibfield  {author} {\bibinfo {author} {\bibfnamefont {M.}~\bibnamefont
			{Evaldsson}}, \bibinfo {author} {\bibfnamefont {I.~V.}\ \bibnamefont
			{Zozoulenko}}, \bibinfo {author} {\bibfnamefont {H.}~\bibnamefont {Xu}}, \
		and\ \bibinfo {author} {\bibfnamefont {T.}~\bibnamefont {Heinzel}},\ }\href
	{https://link.aps.org/doi/10.1103/PhysRevB.78.161407} {\bibfield  {journal}
		{\bibinfo  {journal} {Physical Review B}\ }\textbf {\bibinfo {volume} {78}},\
		\bibinfo {pages} {161407} (\bibinfo {year} {2008})}\BibitemShut {NoStop}%
	\bibitem [{\citenamefont {Baringhaus}\ \emph {et~al.}(2014)\citenamefont
		{Baringhaus}, \citenamefont {Ruan}, \citenamefont {Edler}, \citenamefont
		{Tejeda}, \citenamefont {Sicot}, \citenamefont {Taleb-Ibrahimi},
		\citenamefont {Li}, \citenamefont {Jiang}, \citenamefont {Conrad},
		\citenamefont {Berger}, \citenamefont {Tegenkamp},\ and\ \citenamefont
		{de~Heer}}]{Baringhaus2014}%
	\BibitemOpen
	\bibfield  {author} {\bibinfo {author} {\bibfnamefont {J.}~\bibnamefont
			{Baringhaus}}, \bibinfo {author} {\bibfnamefont {M.}~\bibnamefont {Ruan}},
		\bibinfo {author} {\bibfnamefont {F.}~\bibnamefont {Edler}}, \bibinfo
		{author} {\bibfnamefont {A.}~\bibnamefont {Tejeda}}, \bibinfo {author}
		{\bibfnamefont {M.}~\bibnamefont {Sicot}}, \bibinfo {author} {\bibfnamefont
			{A.}~\bibnamefont {Taleb-Ibrahimi}}, \bibinfo {author} {\bibfnamefont
			{A.-P.}\ \bibnamefont {Li}}, \bibinfo {author} {\bibfnamefont
			{Z.}~\bibnamefont {Jiang}}, \bibinfo {author} {\bibfnamefont {E.~H.}\
			\bibnamefont {Conrad}}, \bibinfo {author} {\bibfnamefont {C.}~\bibnamefont
			{Berger}}, \bibinfo {author} {\bibfnamefont {C.}~\bibnamefont {Tegenkamp}}, \
		and\ \bibinfo {author} {\bibfnamefont {W.~A.}\ \bibnamefont {de~Heer}},\
	}\href {\doibase 10.1038/nature12952} {\bibfield  {journal} {\bibinfo
			{journal} {Nature}\ }\textbf {\bibinfo {volume} {506}},\ \bibinfo {pages}
		{349} (\bibinfo {year} {2014})}\BibitemShut {NoStop}%
	\bibitem [{\citenamefont {Chen}\ \emph {et~al.}(2017)\citenamefont {Chen},
		\citenamefont {He}, \citenamefont {Wang}, \citenamefont {Wang}, \citenamefont
		{Tang}, \citenamefont {Cong}, \citenamefont {Xie}, \citenamefont {Li},
		\citenamefont {Xia}, \citenamefont {Li}, \citenamefont {Wu}, \citenamefont
		{Zhang}, \citenamefont {Deng}, \citenamefont {Yu}, \citenamefont {Xie},\ and\
		\citenamefont {Jiang}}]{Chen2017}%
	\BibitemOpen
	\bibfield  {author} {\bibinfo {author} {\bibfnamefont {L.}~\bibnamefont
			{Chen}}, \bibinfo {author} {\bibfnamefont {L.}~\bibnamefont {He}}, \bibinfo
		{author} {\bibfnamefont {H.~S.}\ \bibnamefont {Wang}}, \bibinfo {author}
		{\bibfnamefont {H.}~\bibnamefont {Wang}}, \bibinfo {author} {\bibfnamefont
			{S.}~\bibnamefont {Tang}}, \bibinfo {author} {\bibfnamefont {C.}~\bibnamefont
			{Cong}}, \bibinfo {author} {\bibfnamefont {H.}~\bibnamefont {Xie}}, \bibinfo
		{author} {\bibfnamefont {L.}~\bibnamefont {Li}}, \bibinfo {author}
		{\bibfnamefont {H.}~\bibnamefont {Xia}}, \bibinfo {author} {\bibfnamefont
			{T.}~\bibnamefont {Li}}, \bibinfo {author} {\bibfnamefont {T.}~\bibnamefont
			{Wu}}, \bibinfo {author} {\bibfnamefont {D.}~\bibnamefont {Zhang}}, \bibinfo
		{author} {\bibfnamefont {L.}~\bibnamefont {Deng}}, \bibinfo {author}
		{\bibfnamefont {T.}~\bibnamefont {Yu}}, \bibinfo {author} {\bibfnamefont
			{X.}~\bibnamefont {Xie}}, \ and\ \bibinfo {author} {\bibfnamefont
			{M.}~\bibnamefont {Jiang}},\ }\href {https://doi.org/10.1038/ncomms14703}
	{\bibfield  {journal} {\bibinfo  {journal} {Nature Communications}\ }\textbf
		{\bibinfo {volume} {8}},\ \bibinfo {pages} {14703} (\bibinfo {year}
		{2017})}\BibitemShut {NoStop}%
	\bibitem [{\citenamefont {Cai}\ \emph {et~al.}(2010)\citenamefont {Cai},
		\citenamefont {Ruffieux}, \citenamefont {Jaafar}, \citenamefont {Bieri},
		\citenamefont {Braun}, \citenamefont {Blankenburg}, \citenamefont {Muoth},
		\citenamefont {Seitsonen}, \citenamefont {Saleh}, \citenamefont {Feng},
		\citenamefont {M\"ullen},\ and\ \citenamefont {Fasel}}]{Cai2010}%
	\BibitemOpen
	\bibfield  {author} {\bibinfo {author} {\bibfnamefont {J.}~\bibnamefont
			{Cai}}, \bibinfo {author} {\bibfnamefont {P.}~\bibnamefont {Ruffieux}},
		\bibinfo {author} {\bibfnamefont {R.}~\bibnamefont {Jaafar}}, \bibinfo
		{author} {\bibfnamefont {M.}~\bibnamefont {Bieri}}, \bibinfo {author}
		{\bibfnamefont {T.}~\bibnamefont {Braun}}, \bibinfo {author} {\bibfnamefont
			{S.}~\bibnamefont {Blankenburg}}, \bibinfo {author} {\bibfnamefont
			{M.}~\bibnamefont {Muoth}}, \bibinfo {author} {\bibfnamefont {A.~P.}\
			\bibnamefont {Seitsonen}}, \bibinfo {author} {\bibfnamefont {M.}~\bibnamefont
			{Saleh}}, \bibinfo {author} {\bibfnamefont {X.}~\bibnamefont {Feng}},
		\bibinfo {author} {\bibfnamefont {K.}~\bibnamefont {M\"ullen}}, \ and\
		\bibinfo {author} {\bibfnamefont {R.}~\bibnamefont {Fasel}},\ }\href
	{http://dx.doi.org/10.1038/nature09211} {\bibfield  {journal} {\bibinfo
			{journal} {Nature}\ }\textbf {\bibinfo {volume} {466}},\ \bibinfo {pages}
		{470} (\bibinfo {year} {2010})}\BibitemShut {NoStop}%
	\bibitem [{\citenamefont {Bennett}\ \emph {et~al.}(2013)\citenamefont
		{Bennett}, \citenamefont {Pedramrazi}, \citenamefont {Madani}, \citenamefont
		{Chen}, \citenamefont {de~Oteyza}, \citenamefont {Chen}, \citenamefont
		{Fischer}, \citenamefont {Crommie},\ and\ \citenamefont
		{Bokor}}]{Bennett2013}%
	\BibitemOpen
	\bibfield  {author} {\bibinfo {author} {\bibfnamefont {P.~B.}\ \bibnamefont
			{Bennett}}, \bibinfo {author} {\bibfnamefont {Z.}~\bibnamefont {Pedramrazi}},
		\bibinfo {author} {\bibfnamefont {A.}~\bibnamefont {Madani}}, \bibinfo
		{author} {\bibfnamefont {Y.-C.}\ \bibnamefont {Chen}}, \bibinfo {author}
		{\bibfnamefont {D.~G.}\ \bibnamefont {de~Oteyza}}, \bibinfo {author}
		{\bibfnamefont {C.}~\bibnamefont {Chen}}, \bibinfo {author} {\bibfnamefont
			{F.~R.}\ \bibnamefont {Fischer}}, \bibinfo {author} {\bibfnamefont {M.~F.}\
			\bibnamefont {Crommie}}, \ and\ \bibinfo {author} {\bibfnamefont
			{J.}~\bibnamefont {Bokor}},\ }\href {\doibase 10.1063/1.4855116} {\bibfield
		{journal} {\bibinfo  {journal} {Applied Physics Letters}\ }\textbf {\bibinfo
			{volume} {103}},\ \bibinfo {pages} {253114} (\bibinfo {year}
		{2013})}\BibitemShut {NoStop}%
	\bibitem [{\citenamefont {Chen}\ \emph {et~al.}(2016)\citenamefont {Chen},
		\citenamefont {Zhang}, \citenamefont {Palma}, \citenamefont {Lodi~Rizzini},
		\citenamefont {Liu}, \citenamefont {Abbas}, \citenamefont {Richter},
		\citenamefont {Martini}, \citenamefont {Wang}, \citenamefont {Cavani},
		\citenamefont {Lu}, \citenamefont {Mishra}, \citenamefont {Coletti},
		\citenamefont {Berger}, \citenamefont {Klappenberger}, \citenamefont
		{Kl\"aui}, \citenamefont {Candini}, \citenamefont {Affronte}, \citenamefont
		{Zhou}, \citenamefont {De~Renzi}, \citenamefont {del Pennino}, \citenamefont
		{Barth}, \citenamefont {R\"ader}, \citenamefont {Narita}, \citenamefont
		{Feng},\ and\ \citenamefont {M\"ullen}}]{Chen2016}%
	\BibitemOpen
	\bibfield  {author} {\bibinfo {author} {\bibfnamefont {Z.}~\bibnamefont
			{Chen}}, \bibinfo {author} {\bibfnamefont {W.}~\bibnamefont {Zhang}},
		\bibinfo {author} {\bibfnamefont {C.-A.}\ \bibnamefont {Palma}}, \bibinfo
		{author} {\bibfnamefont {A.}~\bibnamefont {Lodi~Rizzini}}, \bibinfo {author}
		{\bibfnamefont {B.}~\bibnamefont {Liu}}, \bibinfo {author} {\bibfnamefont
			{A.}~\bibnamefont {Abbas}}, \bibinfo {author} {\bibfnamefont
			{N.}~\bibnamefont {Richter}}, \bibinfo {author} {\bibfnamefont
			{L.}~\bibnamefont {Martini}}, \bibinfo {author} {\bibfnamefont {X.-Y.}\
			\bibnamefont {Wang}}, \bibinfo {author} {\bibfnamefont {N.}~\bibnamefont
			{Cavani}}, \bibinfo {author} {\bibfnamefont {H.}~\bibnamefont {Lu}}, \bibinfo
		{author} {\bibfnamefont {N.}~\bibnamefont {Mishra}}, \bibinfo {author}
		{\bibfnamefont {C.}~\bibnamefont {Coletti}}, \bibinfo {author} {\bibfnamefont
			{R.}~\bibnamefont {Berger}}, \bibinfo {author} {\bibfnamefont
			{F.}~\bibnamefont {Klappenberger}}, \bibinfo {author} {\bibfnamefont
			{M.}~\bibnamefont {Kl\"aui}}, \bibinfo {author} {\bibfnamefont
			{A.}~\bibnamefont {Candini}}, \bibinfo {author} {\bibfnamefont
			{M.}~\bibnamefont {Affronte}}, \bibinfo {author} {\bibfnamefont
			{C.}~\bibnamefont {Zhou}}, \bibinfo {author} {\bibfnamefont {V.}~\bibnamefont
			{De~Renzi}}, \bibinfo {author} {\bibfnamefont {U.}~\bibnamefont {del
				Pennino}}, \bibinfo {author} {\bibfnamefont {J.~V.}\ \bibnamefont {Barth}},
		\bibinfo {author} {\bibfnamefont {H.~J.}\ \bibnamefont {R\"ader}}, \bibinfo
		{author} {\bibfnamefont {A.}~\bibnamefont {Narita}}, \bibinfo {author}
		{\bibfnamefont {X.}~\bibnamefont {Feng}}, \ and\ \bibinfo {author}
		{\bibfnamefont {K.}~\bibnamefont {M\"ullen}},\ }\href {\doibase
		10.1021/jacs.6b10374} {\bibfield  {journal} {\bibinfo  {journal} {Journal of
				the American Chemical Society}\ }\textbf {\bibinfo {volume} {138}},\ \bibinfo
		{pages} {15488} (\bibinfo {year} {2016})}\BibitemShut {NoStop}%
	\bibitem [{\citenamefont {Llinas}\ \emph {et~al.}(2017)\citenamefont {Llinas},
		\citenamefont {Fairbrother}, \citenamefont {Borin~Barin}, \citenamefont
		{Shi}, \citenamefont {Lee}, \citenamefont {Wu}, \citenamefont {Yong~Choi},
		\citenamefont {Braganza}, \citenamefont {Lear}, \citenamefont {Kau},
		\citenamefont {Choi}, \citenamefont {Chen}, \citenamefont {Pedramrazi},
		\citenamefont {Dumslaff}, \citenamefont {Narita}, \citenamefont {Feng},
		\citenamefont {M\"ullen}, \citenamefont {Fischer}, \citenamefont {Zettl},
		\citenamefont {Ruffieux}, \citenamefont {Yablonovitch}, \citenamefont
		{Crommie}, \citenamefont {Fasel},\ and\ \citenamefont {Bokor}}]{Llinas2017}%
	\BibitemOpen
	\bibfield  {author} {\bibinfo {author} {\bibfnamefont {J.~P.}\ \bibnamefont
			{Llinas}}, \bibinfo {author} {\bibfnamefont {A.}~\bibnamefont {Fairbrother}},
		\bibinfo {author} {\bibfnamefont {G.}~\bibnamefont {Borin~Barin}}, \bibinfo
		{author} {\bibfnamefont {W.}~\bibnamefont {Shi}}, \bibinfo {author}
		{\bibfnamefont {K.}~\bibnamefont {Lee}}, \bibinfo {author} {\bibfnamefont
			{S.}~\bibnamefont {Wu}}, \bibinfo {author} {\bibfnamefont {B.}~\bibnamefont
			{Yong~Choi}}, \bibinfo {author} {\bibfnamefont {R.}~\bibnamefont {Braganza}},
		\bibinfo {author} {\bibfnamefont {J.}~\bibnamefont {Lear}}, \bibinfo {author}
		{\bibfnamefont {N.}~\bibnamefont {Kau}}, \bibinfo {author} {\bibfnamefont
			{W.}~\bibnamefont {Choi}}, \bibinfo {author} {\bibfnamefont {C.}~\bibnamefont
			{Chen}}, \bibinfo {author} {\bibfnamefont {Z.}~\bibnamefont {Pedramrazi}},
		\bibinfo {author} {\bibfnamefont {T.}~\bibnamefont {Dumslaff}}, \bibinfo
		{author} {\bibfnamefont {A.}~\bibnamefont {Narita}}, \bibinfo {author}
		{\bibfnamefont {X.}~\bibnamefont {Feng}}, \bibinfo {author} {\bibfnamefont
			{K.}~\bibnamefont {M\"ullen}}, \bibinfo {author} {\bibfnamefont
			{F.}~\bibnamefont {Fischer}}, \bibinfo {author} {\bibfnamefont
			{A.}~\bibnamefont {Zettl}}, \bibinfo {author} {\bibfnamefont
			{P.}~\bibnamefont {Ruffieux}}, \bibinfo {author} {\bibfnamefont
			{E.}~\bibnamefont {Yablonovitch}}, \bibinfo {author} {\bibfnamefont
			{M.}~\bibnamefont {Crommie}}, \bibinfo {author} {\bibfnamefont
			{R.}~\bibnamefont {Fasel}}, \ and\ \bibinfo {author} {\bibfnamefont
			{J.}~\bibnamefont {Bokor}},\ }\href
	{https://doi.org/10.1038/s41467-017-00734-x} {\bibfield  {journal} {\bibinfo
			{journal} {Nature Communications}\ }\textbf {\bibinfo {volume} {8}},\
		\bibinfo {pages} {633} (\bibinfo {year} {2017})}\BibitemShut {NoStop}%
	\bibitem [{\citenamefont {Narita}\ \emph {et~al.}(2014)\citenamefont {Narita},
		\citenamefont {Feng}, \citenamefont {Hernandez}, \citenamefont {Jensen},
		\citenamefont {Bonn}, \citenamefont {Yang}, \citenamefont {Verzhbitskiy},
		\citenamefont {Casiraghi}, \citenamefont {Hansen}, \citenamefont {Koch},
		\citenamefont {Fytas}, \citenamefont {Ivasenko}, \citenamefont {Li},
		\citenamefont {Mali}, \citenamefont {Balandina}, \citenamefont {Mahesh},
		\citenamefont {De~Feyter},\ and\ \citenamefont
		{M\"ullen}}]{Narita2014natchem}%
	\BibitemOpen
	\bibfield  {author} {\bibinfo {author} {\bibfnamefont {A.}~\bibnamefont
			{Narita}}, \bibinfo {author} {\bibfnamefont {X.}~\bibnamefont {Feng}},
		\bibinfo {author} {\bibfnamefont {Y.}~\bibnamefont {Hernandez}}, \bibinfo
		{author} {\bibfnamefont {S.~A.}\ \bibnamefont {Jensen}}, \bibinfo {author}
		{\bibfnamefont {M.}~\bibnamefont {Bonn}}, \bibinfo {author} {\bibfnamefont
			{H.}~\bibnamefont {Yang}}, \bibinfo {author} {\bibfnamefont {I.~A.}\
			\bibnamefont {Verzhbitskiy}}, \bibinfo {author} {\bibfnamefont
			{C.}~\bibnamefont {Casiraghi}}, \bibinfo {author} {\bibfnamefont {M.~R.}\
			\bibnamefont {Hansen}}, \bibinfo {author} {\bibfnamefont {A.~H.~R.}\
			\bibnamefont {Koch}}, \bibinfo {author} {\bibfnamefont {G.}~\bibnamefont
			{Fytas}}, \bibinfo {author} {\bibfnamefont {O.}~\bibnamefont {Ivasenko}},
		\bibinfo {author} {\bibfnamefont {B.}~\bibnamefont {Li}}, \bibinfo {author}
		{\bibfnamefont {K.~S.}\ \bibnamefont {Mali}}, \bibinfo {author}
		{\bibfnamefont {T.}~\bibnamefont {Balandina}}, \bibinfo {author}
		{\bibfnamefont {S.}~\bibnamefont {Mahesh}}, \bibinfo {author} {\bibfnamefont
			{S.}~\bibnamefont {De~Feyter}}, \ and\ \bibinfo {author} {\bibfnamefont
			{K.}~\bibnamefont {M\"ullen}},\ }\href {\doibase 10.1038/NCHEM.1819}
	{\bibfield  {journal} {\bibinfo  {journal} {Nature Chemistry}\ }\textbf
		{\bibinfo {volume} {6}},\ \bibinfo {pages} {126} (\bibinfo {year}
		{2014})}\BibitemShut {NoStop}%
	\bibitem [{\citenamefont {Abbas}\ \emph {et~al.}(2014)\citenamefont {Abbas},
		\citenamefont {Liu}, \citenamefont {Narita}, \citenamefont {Orosco},
		\citenamefont {Feng}, \citenamefont {M\"ullen},\ and\ \citenamefont
		{Zhou}}]{Abbas2014}%
	\BibitemOpen
	\bibfield  {author} {\bibinfo {author} {\bibfnamefont {A.~N.}\ \bibnamefont
			{Abbas}}, \bibinfo {author} {\bibfnamefont {G.}~\bibnamefont {Liu}}, \bibinfo
		{author} {\bibfnamefont {A.}~\bibnamefont {Narita}}, \bibinfo {author}
		{\bibfnamefont {M.}~\bibnamefont {Orosco}}, \bibinfo {author} {\bibfnamefont
			{X.}~\bibnamefont {Feng}}, \bibinfo {author} {\bibfnamefont {K.}~\bibnamefont
			{M\"ullen}}, \ and\ \bibinfo {author} {\bibfnamefont {C.}~\bibnamefont
			{Zhou}},\ }\href {\doibase 10.1021/ja502764d} {\bibfield  {journal} {\bibinfo
			{journal} {Journal of the American Chemical Society}\ }\textbf {\bibinfo
			{volume} {136}},\ \bibinfo {pages} {7555} (\bibinfo {year}
		{2014})}\BibitemShut {NoStop}%
	\bibitem [{\citenamefont {Zschieschang}\ \emph {et~al.}(2015)\citenamefont
		{Zschieschang}, \citenamefont {Klauk}, \citenamefont {M\"uller},
		\citenamefont {Strudwick}, \citenamefont {Hintermann}, \citenamefont
		{Schwab}, \citenamefont {Narita}, \citenamefont {Feng}, \citenamefont
		{M\"ullen},\ and\ \citenamefont {Weitz}}]{Zschieschang2015}%
	\BibitemOpen
	\bibfield  {author} {\bibinfo {author} {\bibfnamefont {U.}~\bibnamefont
			{Zschieschang}}, \bibinfo {author} {\bibfnamefont {H.}~\bibnamefont {Klauk}},
		\bibinfo {author} {\bibfnamefont {I.~B.}\ \bibnamefont {M\"uller}}, \bibinfo
		{author} {\bibfnamefont {A.~J.}\ \bibnamefont {Strudwick}}, \bibinfo {author}
		{\bibfnamefont {T.}~\bibnamefont {Hintermann}}, \bibinfo {author}
		{\bibfnamefont {M.~G.}\ \bibnamefont {Schwab}}, \bibinfo {author}
		{\bibfnamefont {A.}~\bibnamefont {Narita}}, \bibinfo {author} {\bibfnamefont
			{X.}~\bibnamefont {Feng}}, \bibinfo {author} {\bibfnamefont {K.}~\bibnamefont
			{M\"ullen}}, \ and\ \bibinfo {author} {\bibfnamefont {R.~T.}\ \bibnamefont
			{Weitz}},\ }\href {\doibase 10.1002/aelm.201400010} {\bibfield  {journal}
		{\bibinfo  {journal} {Advanced Electronic Materials}\ }\textbf {\bibinfo
			{volume} {1}},\ \bibinfo {pages} {1400010} (\bibinfo {year}
		{2015})}\BibitemShut {NoStop}%
	\bibitem [{\citenamefont {Fantuzzi}\ \emph {et~al.}(2016)\citenamefont
		{Fantuzzi}, \citenamefont {Martini}, \citenamefont {Candini}, \citenamefont
		{Corradini}, \citenamefont {del Pennino}, \citenamefont {Hu}, \citenamefont
		{Feng}, \citenamefont {M\"ullen}, \citenamefont {Narita},\ and\ \citenamefont
		{Affronte}}]{Fantuzzi2016}%
	\BibitemOpen
	\bibfield  {author} {\bibinfo {author} {\bibfnamefont {P.}~\bibnamefont
			{Fantuzzi}}, \bibinfo {author} {\bibfnamefont {L.}~\bibnamefont {Martini}},
		\bibinfo {author} {\bibfnamefont {A.}~\bibnamefont {Candini}}, \bibinfo
		{author} {\bibfnamefont {V.}~\bibnamefont {Corradini}}, \bibinfo {author}
		{\bibfnamefont {U.}~\bibnamefont {del Pennino}}, \bibinfo {author}
		{\bibfnamefont {Y.}~\bibnamefont {Hu}}, \bibinfo {author} {\bibfnamefont
			{X.}~\bibnamefont {Feng}}, \bibinfo {author} {\bibfnamefont {K.}~\bibnamefont
			{M\"ullen}}, \bibinfo {author} {\bibfnamefont {A.}~\bibnamefont {Narita}}, \
		and\ \bibinfo {author} {\bibfnamefont {M.}~\bibnamefont {Affronte}},\ }\href
	{\doibase 10.1016/j.carbon.2016.03.052} {\bibfield  {journal} {\bibinfo
			{journal} {Carbon}\ }\textbf {\bibinfo {volume} {104}},\ \bibinfo {pages}
		{112} (\bibinfo {year} {2016})}\BibitemShut {NoStop}%
	\bibitem [{\citenamefont {Dean}\ \emph {et~al.}(2010)\citenamefont {Dean},
		\citenamefont {Young}, \citenamefont {Meric}, \citenamefont {Lee},
		\citenamefont {Wang}, \citenamefont {Sorgenfrei}, \citenamefont {Watanabe},
		\citenamefont {Taniguchi}, \citenamefont {Kim}, \citenamefont {Shepard},\
		and\ \citenamefont {Hone}}]{Dean2010}%
	\BibitemOpen
	\bibfield  {author} {\bibinfo {author} {\bibfnamefont {C.~R.}\ \bibnamefont
			{Dean}}, \bibinfo {author} {\bibfnamefont {A.~F.}\ \bibnamefont {Young}},
		\bibinfo {author} {\bibfnamefont {I.}~\bibnamefont {Meric}}, \bibinfo
		{author} {\bibfnamefont {C.}~\bibnamefont {Lee}}, \bibinfo {author}
		{\bibfnamefont {L.}~\bibnamefont {Wang}}, \bibinfo {author} {\bibfnamefont
			{S.}~\bibnamefont {Sorgenfrei}}, \bibinfo {author} {\bibfnamefont
			{K.}~\bibnamefont {Watanabe}}, \bibinfo {author} {\bibfnamefont
			{T.}~\bibnamefont {Taniguchi}}, \bibinfo {author} {\bibfnamefont
			{P.}~\bibnamefont {Kim}}, \bibinfo {author} {\bibfnamefont {K.~L.}\
			\bibnamefont {Shepard}}, \ and\ \bibinfo {author} {\bibfnamefont
			{J.}~\bibnamefont {Hone}},\ }\href {\doibase 10.1038/nnano.2010.172}
	{\bibfield  {journal} {\bibinfo  {journal} {Nature Nanotechnology}\ }\textbf
		{\bibinfo {volume} {5}},\ \bibinfo {pages} {722} (\bibinfo {year}
		{2010})}\BibitemShut {NoStop}%
	\bibitem [{\citenamefont {Bischoff}\ \emph {et~al.}(2012)\citenamefont
		{Bischoff}, \citenamefont {Kr\"{a}henmann}, \citenamefont {Dr\"{o}scher},
		\citenamefont {Gruner}, \citenamefont {Barraud}, \citenamefont {Ihn},\ and\
		\citenamefont {Ensslin}}]{Bischoff2012}%
	\BibitemOpen
	\bibfield  {author} {\bibinfo {author} {\bibfnamefont {D.}~\bibnamefont
			{Bischoff}}, \bibinfo {author} {\bibfnamefont {T.}~\bibnamefont
			{Kr\"{a}henmann}}, \bibinfo {author} {\bibfnamefont {S.}~\bibnamefont
			{Dr\"{o}scher}}, \bibinfo {author} {\bibfnamefont {M.~A.}\ \bibnamefont
			{Gruner}}, \bibinfo {author} {\bibfnamefont {C.}~\bibnamefont {Barraud}},
		\bibinfo {author} {\bibfnamefont {T.}~\bibnamefont {Ihn}}, \ and\ \bibinfo
		{author} {\bibfnamefont {K.}~\bibnamefont {Ensslin}},\ }\href {\doibase
		10.1063/1.4765345} {\bibfield  {journal} {\bibinfo  {journal} {Applied
				Physics Letters}\ }\textbf {\bibinfo {volume} {101}},\ \bibinfo {pages}
		{203103} (\bibinfo {year} {2012})}\BibitemShut {NoStop}%
	\bibitem [{\citenamefont {Hu}\ \emph {et~al.}(2018)\citenamefont {Hu},
		\citenamefont {Xie}, \citenamefont {De~Corato}, \citenamefont {Ruini},
		\citenamefont {Zhao}, \citenamefont {Meggendorfer}, \citenamefont {Straas\o},
		\citenamefont {Rondin}, \citenamefont {Simon}, \citenamefont {Li},
		\citenamefont {Finley}, \citenamefont {Hansen}, \citenamefont {Lauret},
		\citenamefont {Molinari}, \citenamefont {Feng}, \citenamefont {Barth},
		\citenamefont {Palma}, \citenamefont {Prezzi}, \citenamefont {M\"ullen},\
		and\ \citenamefont {Narita}}]{Hu2018}%
	\BibitemOpen
	\bibfield  {author} {\bibinfo {author} {\bibfnamefont {Y.}~\bibnamefont
			{Hu}}, \bibinfo {author} {\bibfnamefont {P.}~\bibnamefont {Xie}}, \bibinfo
		{author} {\bibfnamefont {M.}~\bibnamefont {De~Corato}}, \bibinfo {author}
		{\bibfnamefont {A.}~\bibnamefont {Ruini}}, \bibinfo {author} {\bibfnamefont
			{S.}~\bibnamefont {Zhao}}, \bibinfo {author} {\bibfnamefont {F.}~\bibnamefont
			{Meggendorfer}}, \bibinfo {author} {\bibfnamefont {L.~A.}\ \bibnamefont
			{Straas\o}}, \bibinfo {author} {\bibfnamefont {L.}~\bibnamefont {Rondin}},
		\bibinfo {author} {\bibfnamefont {P.}~\bibnamefont {Simon}}, \bibinfo
		{author} {\bibfnamefont {J.}~\bibnamefont {Li}}, \bibinfo {author}
		{\bibfnamefont {J.~J.}\ \bibnamefont {Finley}}, \bibinfo {author}
		{\bibfnamefont {M.~R.}\ \bibnamefont {Hansen}}, \bibinfo {author}
		{\bibfnamefont {J.-S.}\ \bibnamefont {Lauret}}, \bibinfo {author}
		{\bibfnamefont {E.}~\bibnamefont {Molinari}}, \bibinfo {author}
		{\bibfnamefont {X.}~\bibnamefont {Feng}}, \bibinfo {author} {\bibfnamefont
			{J.~V.}\ \bibnamefont {Barth}}, \bibinfo {author} {\bibfnamefont {C.-A.}\
			\bibnamefont {Palma}}, \bibinfo {author} {\bibfnamefont {D.}~\bibnamefont
			{Prezzi}}, \bibinfo {author} {\bibfnamefont {K.}~\bibnamefont {M\"ullen}}, \
		and\ \bibinfo {author} {\bibfnamefont {A.}~\bibnamefont {Narita}},\ }\href
	{\doibase 10.1021/jacs.8b02209} {\bibfield  {journal} {\bibinfo  {journal}
			{J. Am. Chem. Soc.}\ }\textbf {\bibinfo {volume} {140}},\ \bibinfo {pages}
		{7803} (\bibinfo {year} {2018})}\BibitemShut {NoStop}%
	\bibitem [{\citenamefont {Osella}\ \emph {et~al.}(2012)\citenamefont {Osella},
		\citenamefont {Narita}, \citenamefont {Schwab}, \citenamefont {Hernandez},
		\citenamefont {Feng}, \citenamefont {M\"ullen},\ and\ \citenamefont
		{Beljonne}}]{Osella2012}%
	\BibitemOpen
	\bibfield  {author} {\bibinfo {author} {\bibfnamefont {S.}~\bibnamefont
			{Osella}}, \bibinfo {author} {\bibfnamefont {A.}~\bibnamefont {Narita}},
		\bibinfo {author} {\bibfnamefont {M.~G.}\ \bibnamefont {Schwab}}, \bibinfo
		{author} {\bibfnamefont {Y.}~\bibnamefont {Hernandez}}, \bibinfo {author}
		{\bibfnamefont {X.}~\bibnamefont {Feng}}, \bibinfo {author} {\bibfnamefont
			{K.}~\bibnamefont {M\"ullen}}, \ and\ \bibinfo {author} {\bibfnamefont
			{D.}~\bibnamefont {Beljonne}},\ }\href {\doibase 10.1021/nn301478c}
	{\bibfield  {journal} {\bibinfo  {journal} {ACS Nano}\ }\textbf {\bibinfo
			{volume} {6}},\ \bibinfo {pages} {5539} (\bibinfo {year} {2012})}\BibitemShut
	{NoStop}%
	\bibitem [{\citenamefont {Ivanov}\ \emph {et~al.}(2017)\citenamefont {Ivanov},
		\citenamefont {Hu}, \citenamefont {Osella}, \citenamefont {Beser},
		\citenamefont {Wang}, \citenamefont {Beljonne}, \citenamefont {Narita},
		\citenamefont {M\"ullen}, \citenamefont {Turchinovich},\ and\ \citenamefont
		{Bonn}}]{Ivanov2017}%
	\BibitemOpen
	\bibfield  {author} {\bibinfo {author} {\bibfnamefont {I.}~\bibnamefont
			{Ivanov}}, \bibinfo {author} {\bibfnamefont {Y.}~\bibnamefont {Hu}}, \bibinfo
		{author} {\bibfnamefont {S.}~\bibnamefont {Osella}}, \bibinfo {author}
		{\bibfnamefont {U.}~\bibnamefont {Beser}}, \bibinfo {author} {\bibfnamefont
			{H.~I.}\ \bibnamefont {Wang}}, \bibinfo {author} {\bibfnamefont
			{D.}~\bibnamefont {Beljonne}}, \bibinfo {author} {\bibfnamefont
			{A.}~\bibnamefont {Narita}}, \bibinfo {author} {\bibfnamefont
			{K.}~\bibnamefont {M\"ullen}}, \bibinfo {author} {\bibfnamefont
			{D.}~\bibnamefont {Turchinovich}}, \ and\ \bibinfo {author} {\bibfnamefont
			{M.}~\bibnamefont {Bonn}},\ }\href {\doibase 10.1021/jacs.7b03467} {\bibfield
		{journal} {\bibinfo  {journal} {J. Am. Chem. Soc.}\ }\textbf {\bibinfo
			{volume} {139}},\ \bibinfo {pages} {7982} (\bibinfo {year}
		{2017})}\BibitemShut {NoStop}%
	\bibitem [{\citenamefont {Villegas}\ \emph {et~al.}(2014)\citenamefont
		{Villegas}, \citenamefont {Mendon\c{c}a},\ and\ \citenamefont
		{Rocha}}]{Villegas2014}%
	\BibitemOpen
	\bibfield  {author} {\bibinfo {author} {\bibfnamefont {C.~E.~P.}\
			\bibnamefont {Villegas}}, \bibinfo {author} {\bibfnamefont {P.~B.}\
			\bibnamefont {Mendon\c{c}a}}, \ and\ \bibinfo {author} {\bibfnamefont
			{A.~R.}\ \bibnamefont {Rocha}},\ }\href {\doibase 10.1038/srep06579}
	{\bibfield  {journal} {\bibinfo  {journal} {Scientific Reports}\ }\textbf
		{\bibinfo {volume} {4}},\ \bibinfo {pages} {6579} (\bibinfo {year}
		{2014})}\BibitemShut {NoStop}%
	\bibitem [{\citenamefont {Ouyang}\ \emph {et~al.}(2018)\citenamefont {Ouyang},
		\citenamefont {Mandelli}, \citenamefont {Urbakh},\ and\ \citenamefont
		{Hod}}]{Ouyang2018}%
	\BibitemOpen
	\bibfield  {author} {\bibinfo {author} {\bibfnamefont {W.}~\bibnamefont
			{Ouyang}}, \bibinfo {author} {\bibfnamefont {D.}~\bibnamefont {Mandelli}},
		\bibinfo {author} {\bibfnamefont {M.}~\bibnamefont {Urbakh}}, \ and\ \bibinfo
		{author} {\bibfnamefont {O.}~\bibnamefont {Hod}},\ }\href {\doibase
		10.1021/acs.nanolett.8b02848} {\bibfield  {journal} {\bibinfo  {journal}
			{Nano Letters}\ }\textbf {\bibinfo {volume} {18}},\ \bibinfo {pages} {6009}
		(\bibinfo {year} {2018})}\BibitemShut {NoStop}%
	\bibitem [{\citenamefont {Kawai}\ \emph {et~al.}(2016)\citenamefont {Kawai},
		\citenamefont {Benassi}, \citenamefont {Gnecco}, \citenamefont {S\"ode},
		\citenamefont {Pawlak}, \citenamefont {Feng}, \citenamefont {M\"ullen},
		\citenamefont {Passerone}, \citenamefont {Pignedoli}, \citenamefont
		{Ruffieux}, \citenamefont {Fasel},\ and\ \citenamefont {Meyer}}]{Kawai2016}%
	\BibitemOpen
	\bibfield  {author} {\bibinfo {author} {\bibfnamefont {S.}~\bibnamefont
			{Kawai}}, \bibinfo {author} {\bibfnamefont {A.}~\bibnamefont {Benassi}},
		\bibinfo {author} {\bibfnamefont {E.}~\bibnamefont {Gnecco}}, \bibinfo
		{author} {\bibfnamefont {H.}~\bibnamefont {S\"ode}}, \bibinfo {author}
		{\bibfnamefont {R.}~\bibnamefont {Pawlak}}, \bibinfo {author} {\bibfnamefont
			{X.}~\bibnamefont {Feng}}, \bibinfo {author} {\bibfnamefont {K.}~\bibnamefont
			{M\"ullen}}, \bibinfo {author} {\bibfnamefont {D.}~\bibnamefont {Passerone}},
		\bibinfo {author} {\bibfnamefont {C.~A.}\ \bibnamefont {Pignedoli}}, \bibinfo
		{author} {\bibfnamefont {P.}~\bibnamefont {Ruffieux}}, \bibinfo {author}
		{\bibfnamefont {R.}~\bibnamefont {Fasel}}, \ and\ \bibinfo {author}
		{\bibfnamefont {E.}~\bibnamefont {Meyer}},\ }\href {\doibase
		10.1126/science.aad3569} {\bibfield  {journal} {\bibinfo  {journal}
			{Science}\ }\textbf {\bibinfo {volume} {351}},\ \bibinfo {pages} {957}
		(\bibinfo {year} {2016})}\BibitemShut {NoStop}%
	\bibitem [{\citenamefont {Taniguchi}\ and\ \citenamefont
		{Watanabe}(2007)}]{Taniguchi2007}%
	\BibitemOpen
	\bibfield  {author} {\bibinfo {author} {\bibfnamefont {T.}~\bibnamefont
			{Taniguchi}}\ and\ \bibinfo {author} {\bibfnamefont {K.}~\bibnamefont
			{Watanabe}},\ }\href {\doibase 10.1016/j.jcrysgro.2006.12.061} {\bibfield
		{journal} {\bibinfo  {journal} {Journal of Crystal Growth}\ }\textbf
		{\bibinfo {volume} {303}},\ \bibinfo {pages} {525} (\bibinfo {year}
		{2007})}\BibitemShut {NoStop}%
	\bibitem [{\citenamefont {Xia}\ \emph {et~al.}(2011)\citenamefont {Xia},
		\citenamefont {Perebeinos}, \citenamefont {Lin}, \citenamefont {Wu},\ and\
		\citenamefont {Avouris}}]{Xia2011}%
	\BibitemOpen
	\bibfield  {author} {\bibinfo {author} {\bibfnamefont {F.}~\bibnamefont
			{Xia}}, \bibinfo {author} {\bibfnamefont {V.}~\bibnamefont {Perebeinos}},
		\bibinfo {author} {\bibfnamefont {Y.-m.}\ \bibnamefont {Lin}}, \bibinfo
		{author} {\bibfnamefont {Y.}~\bibnamefont {Wu}}, \ and\ \bibinfo {author}
		{\bibfnamefont {P.}~\bibnamefont {Avouris}},\ }\href
	{https://doi.org/10.1038/nnano.2011.6} {\bibfield  {journal} {\bibinfo
			{journal} {Nature Nanotechnology}\ }\textbf {\bibinfo {volume} {6}},\
		\bibinfo {pages} {179} (\bibinfo {year} {2011})}\BibitemShut {NoStop}%
	\bibitem [{\citenamefont {Perello}\ \emph {et~al.}(2011)\citenamefont
		{Perello}, \citenamefont {Lim}, \citenamefont {Chae}, \citenamefont {Lee},
		\citenamefont {Kim}, \citenamefont {Lee},\ and\ \citenamefont
		{Yun}}]{Perello2011}%
	\BibitemOpen
	\bibfield  {author} {\bibinfo {author} {\bibfnamefont {D.~J.}\ \bibnamefont
			{Perello}}, \bibinfo {author} {\bibfnamefont {S.~C.}\ \bibnamefont {Lim}},
		\bibinfo {author} {\bibfnamefont {S.~J.}\ \bibnamefont {Chae}}, \bibinfo
		{author} {\bibfnamefont {I.}~\bibnamefont {Lee}}, \bibinfo {author}
		{\bibfnamefont {M.~J.}\ \bibnamefont {Kim}}, \bibinfo {author} {\bibfnamefont
			{Y.~H.}\ \bibnamefont {Lee}}, \ and\ \bibinfo {author} {\bibfnamefont
			{M.}~\bibnamefont {Yun}},\ }\href {\doibase 10.1021/nn102343k} {\bibfield
		{journal} {\bibinfo  {journal} {ACS Nano}\ }\textbf {\bibinfo {volume} {5}},\
		\bibinfo {pages} {1756} (\bibinfo {year} {2011})}\BibitemShut {NoStop}%
	\bibitem [{\citenamefont {Zhang}\ \emph {et~al.}(2007)\citenamefont {Zhang},
		\citenamefont {Yao}, \citenamefont {Liu}, \citenamefont {Jin}, \citenamefont
		{Liang}, \citenamefont {Chen},\ and\ \citenamefont {Peng}}]{Zhang2007}%
	\BibitemOpen
	\bibfield  {author} {\bibinfo {author} {\bibfnamefont {Z.}~\bibnamefont
			{Zhang}}, \bibinfo {author} {\bibfnamefont {K.}~\bibnamefont {Yao}}, \bibinfo
		{author} {\bibfnamefont {Y.}~\bibnamefont {Liu}}, \bibinfo {author}
		{\bibfnamefont {C.}~\bibnamefont {Jin}}, \bibinfo {author} {\bibfnamefont
			{X.}~\bibnamefont {Liang}}, \bibinfo {author} {\bibfnamefont
			{Q.}~\bibnamefont {Chen}}, \ and\ \bibinfo {author} {\bibfnamefont {L.-M.}\
			\bibnamefont {Peng}},\ }\href {\doibase 10.1002/adfm.200600475} {\bibfield
		{journal} {\bibinfo  {journal} {Adv. Funct. Mater.}\ }\textbf {\bibinfo
			{volume} {17}},\ \bibinfo {pages} {2478} (\bibinfo {year}
		{2007})}\BibitemShut {NoStop}%
	\bibitem [{\citenamefont {Appenzeller}\ \emph {et~al.}(2002)\citenamefont
		{Appenzeller}, \citenamefont {Knoch}, \citenamefont {Derycke}, \citenamefont
		{Martel}, \citenamefont {Wind},\ and\ \citenamefont
		{Avouris}}]{Appenzeller2002}%
	\BibitemOpen
	\bibfield  {author} {\bibinfo {author} {\bibfnamefont {J.}~\bibnamefont
			{Appenzeller}}, \bibinfo {author} {\bibfnamefont {J.}~\bibnamefont {Knoch}},
		\bibinfo {author} {\bibfnamefont {V.}~\bibnamefont {Derycke}}, \bibinfo
		{author} {\bibfnamefont {R.}~\bibnamefont {Martel}}, \bibinfo {author}
		{\bibfnamefont {S.}~\bibnamefont {Wind}}, \ and\ \bibinfo {author}
		{\bibfnamefont {P.}~\bibnamefont {Avouris}},\ }\href {\doibase
		10.1103/PhysRevLett.89.126801} {\bibfield  {journal} {\bibinfo  {journal}
			{Phys. Rev. Lett.}\ }\textbf {\bibinfo {volume} {89}},\ \bibinfo {pages}
		{126801} (\bibinfo {year} {2002})}\BibitemShut {NoStop}%
	\bibitem [{\citenamefont {Chiquito}\ \emph {et~al.}(2012)\citenamefont
		{Chiquito}, \citenamefont {Amorim}, \citenamefont {Berengue}, \citenamefont
		{Araujo}, \citenamefont {Bernardo},\ and\ \citenamefont
		{Leite}}]{Chiquito2012}%
	\BibitemOpen
	\bibfield  {author} {\bibinfo {author} {\bibfnamefont {A.~J.}\ \bibnamefont
			{Chiquito}}, \bibinfo {author} {\bibfnamefont {C.~A.}\ \bibnamefont
			{Amorim}}, \bibinfo {author} {\bibfnamefont {O.~M.}\ \bibnamefont
			{Berengue}}, \bibinfo {author} {\bibfnamefont {L.~S.}\ \bibnamefont
			{Araujo}}, \bibinfo {author} {\bibfnamefont {E.~P.}\ \bibnamefont
			{Bernardo}}, \ and\ \bibinfo {author} {\bibfnamefont {E.~R.}\ \bibnamefont
			{Leite}},\ }\href {http://stacks.iop.org/0953-8984/24/i=22/a=225303}
	{\bibfield  {journal} {\bibinfo  {journal} {Journal of Physics: Condensed
				Matter}\ }\textbf {\bibinfo {volume} {24}},\ \bibinfo {pages} {225303}
		(\bibinfo {year} {2012})}\BibitemShut {NoStop}%
	\bibitem [{\citenamefont {Sze}\ and\ \citenamefont {Ng}(2007)}]{Sze2007}%
	\BibitemOpen
	\bibfield  {author} {\bibinfo {author} {\bibfnamefont {S.~M.}\ \bibnamefont
			{Sze}}\ and\ \bibinfo {author} {\bibfnamefont {K.~K.}\ \bibnamefont {Ng}},\
	}\href@noop {} {\emph {\bibinfo {title} {Physics of Semiconductor
				Devices}}},\ \bibinfo {edition} {3rd}\ ed.\ (\bibinfo  {publisher} {John
		Wiley \& Sons},\ \bibinfo {year} {2007})\BibitemShut {NoStop}%
	\bibitem [{\citenamefont {Padovani}\ and\ \citenamefont
		{Stratton}(1966)}]{Padovani1966}%
	\BibitemOpen
	\bibfield  {author} {\bibinfo {author} {\bibfnamefont {F.~A.}\ \bibnamefont
			{Padovani}}\ and\ \bibinfo {author} {\bibfnamefont {R.}~\bibnamefont
			{Stratton}},\ }\href@noop {} {\bibfield  {journal} {\bibinfo  {journal}
			{Solid-State Electronics}\ }\textbf {\bibinfo {volume} {9}},\ \bibinfo
		{pages} {695} (\bibinfo {year} {1966})}\BibitemShut {NoStop}%
	\bibitem [{\citenamefont {Liu}\ \emph {et~al.}(2008)\citenamefont {Liu},
		\citenamefont {Zhang}, \citenamefont {Hu}, \citenamefont {Jin},\ and\
		\citenamefont {Peng}}]{Liu2008}%
	\BibitemOpen
	\bibfield  {author} {\bibinfo {author} {\bibfnamefont {Y.}~\bibnamefont
			{Liu}}, \bibinfo {author} {\bibfnamefont {Z.~Y.}\ \bibnamefont {Zhang}},
		\bibinfo {author} {\bibfnamefont {Y.~F.}\ \bibnamefont {Hu}}, \bibinfo
		{author} {\bibfnamefont {C.~H.}\ \bibnamefont {Jin}}, \ and\ \bibinfo
		{author} {\bibfnamefont {L.-M.}\ \bibnamefont {Peng}},\ }\href@noop {}
	{\bibfield  {journal} {\bibinfo  {journal} {J. Nanosc. Nanotechnol.}\
		}\textbf {\bibinfo {volume} {8}},\ \bibinfo {pages} {252} (\bibinfo {year}
		{2008})}\BibitemShut {NoStop}%
	\bibitem [{\citenamefont {Crowell}(1965)}]{Crowell1965}%
	\BibitemOpen
	\bibfield  {author} {\bibinfo {author} {\bibfnamefont {C.}~\bibnamefont
			{Crowell}},\ }\href {\doibase https://doi.org/10.1016/0038-1101(65)90116-4}
	{\bibfield  {journal} {\bibinfo  {journal} {Solid-State Electronics}\
		}\textbf {\bibinfo {volume} {8}},\ \bibinfo {pages} {395 } (\bibinfo {year}
		{1965})}\BibitemShut {NoStop}%
	\bibitem [{\citenamefont {Svensson}\ and\ \citenamefont
		{Campbell}(2011)}]{Svensson2011}%
	\BibitemOpen
	\bibfield  {author} {\bibinfo {author} {\bibfnamefont {J.}~\bibnamefont
			{Svensson}}\ and\ \bibinfo {author} {\bibfnamefont {E.~E.~B.}\ \bibnamefont
			{Campbell}},\ }\href {\doibase 10.1063/1.3664139} {\bibfield  {journal}
		{\bibinfo  {journal} {Journal of Applied Physics}\ }\textbf {\bibinfo
			{volume} {110}},\ \bibinfo {pages} {111101} (\bibinfo {year}
		{2011})}\BibitemShut {NoStop}%
	\bibitem [{\citenamefont {Konnerth}\ \emph {et~al.}(2015)\citenamefont
		{Konnerth}, \citenamefont {Cervetti}, \citenamefont {Narita}, \citenamefont
		{Feng}, \citenamefont {M\"ullen}, \citenamefont {Hoyer}, \citenamefont
		{Burghard}, \citenamefont {Kern}, \citenamefont {Dressel},\ and\
		\citenamefont {Bogani}}]{Konnerth2015}%
	\BibitemOpen
	\bibfield  {author} {\bibinfo {author} {\bibfnamefont {R.}~\bibnamefont
			{Konnerth}}, \bibinfo {author} {\bibfnamefont {C.}~\bibnamefont {Cervetti}},
		\bibinfo {author} {\bibfnamefont {A.}~\bibnamefont {Narita}}, \bibinfo
		{author} {\bibfnamefont {X.}~\bibnamefont {Feng}}, \bibinfo {author}
		{\bibfnamefont {K.}~\bibnamefont {M\"ullen}}, \bibinfo {author}
		{\bibfnamefont {A.}~\bibnamefont {Hoyer}}, \bibinfo {author} {\bibfnamefont
			{M.}~\bibnamefont {Burghard}}, \bibinfo {author} {\bibfnamefont
			{K.}~\bibnamefont {Kern}}, \bibinfo {author} {\bibfnamefont {M.}~\bibnamefont
			{Dressel}}, \ and\ \bibinfo {author} {\bibfnamefont {L.}~\bibnamefont
			{Bogani}},\ }\href {\doibase 10.1039/c4nr07378a} {\bibfield  {journal}
		{\bibinfo  {journal} {Nanoscale}\ }\textbf {\bibinfo {volume} {7}},\ \bibinfo
		{pages} {12807} (\bibinfo {year} {2015})}\BibitemShut {NoStop}%
	\bibitem [{\citenamefont {Matsuda}\ \emph {et~al.}(2010)\citenamefont
		{Matsuda}, \citenamefont {Deng},\ and\ \citenamefont
		{Goddard}}]{Matsuda2010}%
	\BibitemOpen
	\bibfield  {author} {\bibinfo {author} {\bibfnamefont {Y.}~\bibnamefont
			{Matsuda}}, \bibinfo {author} {\bibfnamefont {W.-Q.}\ \bibnamefont {Deng}}, \
		and\ \bibinfo {author} {\bibfnamefont {W.~A.}\ \bibnamefont {Goddard}},\
	}\href {\doibase 10.1021/jp806437y} {\bibfield  {journal} {\bibinfo
			{journal} {J. Phys. Chem. C}\ }\textbf {\bibinfo {volume} {114}},\ \bibinfo
		{pages} {17845} (\bibinfo {year} {2010})}\BibitemShut {NoStop}%
	\bibitem [{\citenamefont {Wang}\ \emph {et~al.}(2013)\citenamefont {Wang},
		\citenamefont {Meric}, \citenamefont {Huang}, \citenamefont {Gao},
		\citenamefont {Gao}, \citenamefont {Tran}, \citenamefont {Taniguchi},
		\citenamefont {Watanabe}, \citenamefont {Campos}, \citenamefont {Muller},
		\citenamefont {Guo}, \citenamefont {Kim}, \citenamefont {Hone}, \citenamefont
		{Shepard},\ and\ \citenamefont {Dean}}]{Wang1Dcontacts}%
	\BibitemOpen
	\bibfield  {author} {\bibinfo {author} {\bibfnamefont {L.}~\bibnamefont
			{Wang}}, \bibinfo {author} {\bibfnamefont {I.}~\bibnamefont {Meric}},
		\bibinfo {author} {\bibfnamefont {P.~Y.}\ \bibnamefont {Huang}}, \bibinfo
		{author} {\bibfnamefont {Q.}~\bibnamefont {Gao}}, \bibinfo {author}
		{\bibfnamefont {Y.}~\bibnamefont {Gao}}, \bibinfo {author} {\bibfnamefont
			{H.}~\bibnamefont {Tran}}, \bibinfo {author} {\bibfnamefont {T.}~\bibnamefont
			{Taniguchi}}, \bibinfo {author} {\bibfnamefont {K.}~\bibnamefont {Watanabe}},
		\bibinfo {author} {\bibfnamefont {L.~M.}\ \bibnamefont {Campos}}, \bibinfo
		{author} {\bibfnamefont {D.~A.}\ \bibnamefont {Muller}}, \bibinfo {author}
		{\bibfnamefont {J.}~\bibnamefont {Guo}}, \bibinfo {author} {\bibfnamefont
			{P.}~\bibnamefont {Kim}}, \bibinfo {author} {\bibfnamefont {J.}~\bibnamefont
			{Hone}}, \bibinfo {author} {\bibfnamefont {K.~L.}\ \bibnamefont {Shepard}}, \
		and\ \bibinfo {author} {\bibfnamefont {C.~R.}\ \bibnamefont {Dean}},\ }\href
	{\doibase 10.1126/science.1244358} {\bibfield  {journal} {\bibinfo  {journal}
			{Science}\ }\textbf {\bibinfo {volume} {342}},\ \bibinfo {pages} {614}
		(\bibinfo {year} {2013})}\BibitemShut {NoStop}%
	\bibitem [{\citenamefont {Huang}\ \emph {et~al.}(2015)\citenamefont {Huang},
		\citenamefont {Pan}, \citenamefont {Tran}, \citenamefont {Cheng},
		\citenamefont {Watanabe}, \citenamefont {Taniguchi}, \citenamefont {Lau},\
		and\ \citenamefont {Bockrath}}]{Huang20150D}%
	\BibitemOpen
	\bibfield  {author} {\bibinfo {author} {\bibfnamefont {J.-W.}\ \bibnamefont
			{Huang}}, \bibinfo {author} {\bibfnamefont {C.}~\bibnamefont {Pan}}, \bibinfo
		{author} {\bibfnamefont {S.}~\bibnamefont {Tran}}, \bibinfo {author}
		{\bibfnamefont {B.}~\bibnamefont {Cheng}}, \bibinfo {author} {\bibfnamefont
			{K.}~\bibnamefont {Watanabe}}, \bibinfo {author} {\bibfnamefont
			{T.}~\bibnamefont {Taniguchi}}, \bibinfo {author} {\bibfnamefont {C.~N.}\
			\bibnamefont {Lau}}, \ and\ \bibinfo {author} {\bibfnamefont
			{M.}~\bibnamefont {Bockrath}},\ }\href {\doibase
		10.1021/acs.nanolett.5b02716} {\bibfield  {journal} {\bibinfo  {journal}
			{Nano Letters}\ }\textbf {\bibinfo {volume} {15}},\ \bibinfo {pages} {6836}
		(\bibinfo {year} {2015})}\BibitemShut {NoStop}%
\end{thebibliography}




\section*{Supplementary material}

# Graphene Nanoribbons on Hexagonal Boron Nitride: Deposition and Transport Characterization – Supplementary Material

Tobias Preis, Christian Kick, Andreas Lex, Dieter Weiss, and Jonathan Eroms
*Institute of Experimental and Applied Physics,*
*University of Regensburg, D-93040 Regensburg, Germany*

Akimitsu Narita, Yunbin Hu,[*] and Klaus Müllen
*Max Planck Institute for Polymer Research,*
*Ackermannweg 10, D-55128 Mainz, Germany*

Kenji Watanabe and Takashi Taniguchi
*National Institute for Materials Science,*
*1-1 Namiki, Tsukuba 305-0044, Japan*

## I. Absence of adsorption to SiO$_2$

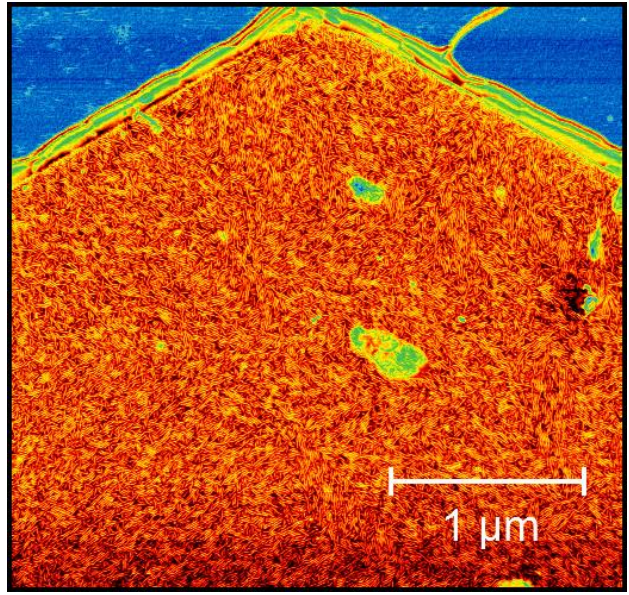

FIG. S1: AFM phase image of an hBN flake (orange region) exfoliated on SiO$_2$ (blue region). Here, a dense array of cGNRs on the hBN is clearly seen. In contrast, the SiO$_2$ shows almost no signs of adsorbed cGNRs. The color scale was optimized for optimum contrast in both regions.

While on hBN flakes we regularly observe densely packed arrays of cGNRs, on the SiO$_2$ substrate adsorption is suppressed. In the AFM image shown in Fig. S1, this is clearly visible.

## II. Agglomeration of cGNRs after annealing

To improve sample cleanliness, we also tried annealing the samples after deposition of cGNRs onto the hBN flakes. This was performed in a semiconductor annealing oven in a low pressure atmosphere (ramp up at 8 mbar, annealing at 8 mbar and finally 100 mbar) of forming gas, at a temperature of 450°, for 2 hours. As can be seen in Fig. S2, the cGNRs change their morphology and form agglomerates. Therefore, we refrained from annealing the transport devices.

## III. Length distribution of graphene nanoribbons

We evaluated the length distribution from both 4-cGNRs and 6-cGNRs after deposition onto hBN from an AFM image. The resulting histograms are shown in Fig. S3. For the 4-cGNRs we find an average length of 90 nm, with standard deviation 60 nm, and for the 6-cGNRs we find an average length of 100 nm, with standard deviation 70 nm.

## IV. Models for $I$-$V$ characteristic of a Schottky barrier and details on the fitting

In the model of thermionic emission (Schottky characteristic), the current density $J$ through a metal-semiconductor contact is given by:[1]

$$J = AT^2 \exp\left(-\frac{\Phi}{kT}\right) \exp\left(\frac{qV}{nkT}\right) \left[1 - \exp\left(-\frac{qV}{kT}\right)\right] \tag{S1}$$

Here, $A$ is the Richardson constant, $T$ the temperature, $\Phi$ the height of the Schottky barrier, $k$ the Boltzmann constant, $q$ the charge, $V$ the voltage applied across the barrier and $n$ the ideality factor, which describes how the barrier height changes when applying a voltage across it. Here, only thermal activation over the Schottky barrier is considered. For an ideal junction $n = 1$, the

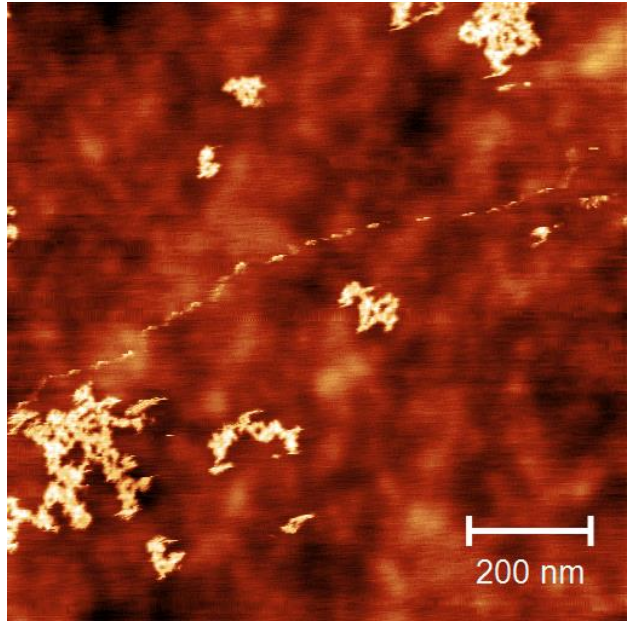

FIG. S2: AFM topography image of a cGNR covered hBN flake after annealing to 450° for 2 hours. The cGNRs apparently have formed agglomerates and the bare hBN surface is visible. The high mobility of cGNRs on the hBN surface at elevated temperatures could be explained by the atomic flatness of the hBN substrate.

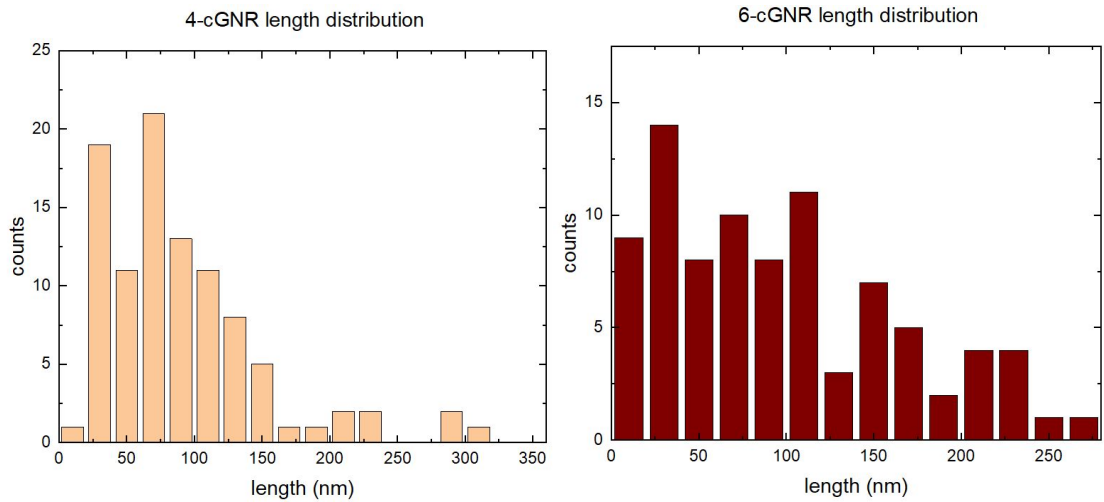

FIG. S3: Length distributions of 4-cGNRs (left) and 6-cGNRs (right) after sonication and deposition on hBN, measured by AFM.

current under large reverse bias is constant. As in our case the device can be considered as back to back Schottky diodes, the junction at reverse bias dominates the total current. Since this is seen to increase with bias, $n$ must be larger than one. Plotting the data on a semi-log scale, the reverse current in the thermionic emission model yields a straight line, which gives a poor description of the experimental data (see Fig. S4).

3

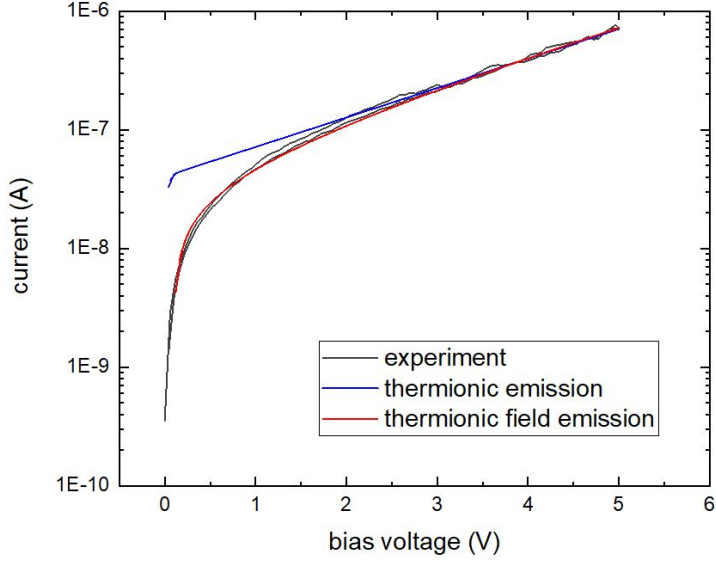

FIG. S4: Comparison of thermionic emission model (blue) and thermionic field emission model on a semi-log scale for a Pd 6-cGNR sample.

If electron tunneling at elevated temperatures is considered, electrons are thermally activated to a higher energy, still below the barrier, where the tunneling probability is enhanced. This is described in the thermionic field emission model[2–5] (Eq. (1) of the main text, which is reproduced here for convenience):

$$ J = \frac{AT\sqrt{\pi q E_{00}}}{k} \exp\left(-\frac{\Phi}{qE_0}\right) \exp\left[V\left(\frac{q}{kT}-\frac{1}{E_0}\right)\right] \times \sqrt{q(V-\zeta) + \frac{\Phi}{\cosh^2(qE_{00}/kT)}} $$

Here, $A$ is the Richardson constant, $T$ the temperature, $\Phi$ the height of the Schottky barrier, $k$ the Boltzmann constant, $q$ the elementary charge, $V > 0$ the voltage applied across the barrier, and $\zeta$ the distance between the band edge and the Fermi level. $E_{00}$ describes the shape of the barrier[2] and $E_0 = E_{00}\coth(qE_{00}/kT)$. In the context of Schottky barriers, $E_{00}$ is usually given in terms of parameters for a metal-doped semiconductor junction[1], but originally stems from a Taylor expansion of the barrier shape[2]. It can therefore also be applied to our situation. The formula for $J$ is valid in reverse direction, for a bias larger than $qV > 3kT$. As can be seen, the semi-log plot now gives a much more faithful reproduction of the experimental data.

For the contact area, we take the cross-sectional area to be the width of a nanoribbon times the layer distance in graphene, $i.e.$ 0.24 nm$^2$ for the 4-cGNRs and 0.39 nm$^2$ for the 6-cGNRs and fit the data using this number. The number $N$ of ribbons in parallel is estimated in the following way. From the AFM images, we find a length distribution $P(L)$ shown in Fig. S3. If the electrode spacing is $d < L$, the probability to hit a single ribbon of length $L$ is $p = (L-d)/L$. The electrodes are aligned in such a way that the ribbon direction in one domain is perpendicular to the contact electrodes, meaning an optimum alignment. As all three domain types occur with equal probability, the electrode spacing is $d$ for one third of the electrode width, and effectively $d/\cos(60°)$ for two thirds of the electrode width. Further, we evaluate the mean distance between GNRs from the

4

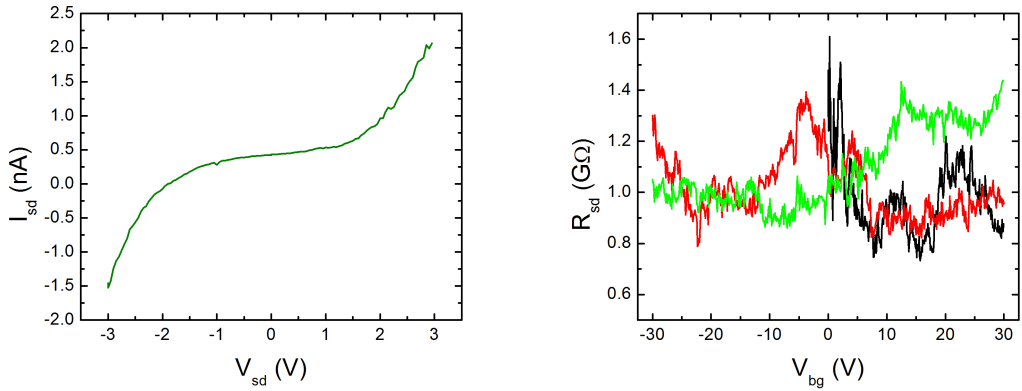

FIG. S5: Left: *I-V* characteristic of a 4-cGNR sample with Pd contacts at zero gate voltage. Right: Gate response of the source-drain resistance at fixed bias. There is no clear dependence on gate voltage in several runs.

AFM images, and also take into account the comb-like contact geometry. From all this, we arrive at a maximum number of parallel nanoribbons of $N = 1000$ for the 4-cGNR, and $N = 300$ for the 6-cGNR samples. This is the dominant error, exceeding further sources, such as the uncertainty in the effective mass entering into the Richardson constant $A$ and a possible barrier between the metal and the GNR, diminishing the current. Due to the exponential dependence on the barrier height, this amounts to an uncertainty in $\Phi$ of 150 meV for the 6-cGNRs and 180 meV for the 4-cGNRs. We believe that the uncertainty in the number of parallel ribbons is largely responsible for the large difference in observed currents for comparable devices.

## V. Back gate response of the sample resistance

In addition to the gate-dependent $I_d$-$V_{sd}$ curves shown in the main text, here we include additional data from a different device. In Fig. S5 we see consecutive $R_{sd}$-$V_{bg}$ curves taken at a fixed source-drain bias voltage of $V_{sd} = 2.5$ V. As can be seen, there is no clear gate dependence, but the sample resistance shows run-to-run variations. The lack of a gate dependence is in accordance to the $I_d$-$V_{sd}$ curves shown in the main text, where curves taken at different gate voltages are identical. Therefore, it is not possible to extract information such as the background doping, or carrier mobility.

---

\* Current address:Department of Organic and Polymer Chemistry, College of Chemistry and Chemical Engineering, Central South University, Changsha, Hunan 410083, China

[1] S. M. Sze and K. K. Ng, *Physics of Semiconductor Devices*, 3rd ed. (John Wiley & Sons, 2007).

[2] F. A. Padovani and R. Stratton, Solid-State Electronics **9**, 695 (1966).

[3] Z. Zhang, K. Yao, Y. Liu, C. Jin, X. Liang, Q. Chen, and L.-M. Peng, Adv. Funct. Mater. **17**, 2478 (2007).

[4] Y. Liu, Z. Y. Zhang, Y. F. Hu, C. H. Jin, and L.-M. Peng, J. Nanosc. Nanotechnol. **8**, 252 (2008).

[5] D. J. Perello, S. C. Lim, S. J. Chae, I. Lee, M. J. Kim, Y. H. Lee, and M. Yun, ACS Nano **5**, 1756 (2011).



\end{document}